\documentclass{aa}

\usepackage{mhchem}
\usepackage{graphicx}
\usepackage{txfonts}
\usepackage{lipsum}
\usepackage{subcaption}
\usepackage{lscape}
\usepackage{placeins}

\usepackage[hidelinks]{hyperref}

\begin{document}

   \title{Origins of complex organic molecules in L1551 IRS 5}
   \subtitle{Interplay of thermal heating and accretion shocks}
\author{Ji-hyun Kang\inst{1}
    \and Jeong-Eun Lee\inst{2}
    \and Seokho Lee\inst{1}
    \and Shigehisa Takakuwa\inst{3,4}
}

\institute{
    Korea Astronomy and Space Science Institute, 776 Daedeokdae-ro, Yuseong-gu, Daejeon 34055, Republic of Korea\\
    \email{jkang@kasi.re.kr}
    \and
    Department of Physics and Astronomy, Seoul National University, 1 Gwanak-ro, Gwanak-gu, Seoul 08826, Republic of Korea
    \and
    Department of Physics and Astronomy, Graduate School of Science and Engineering, Kagoshima University, 1-21-35 Korimoto, Kagoshima, Kagoshima 890-0065, Japan
    \and
    Academia Sinica Institute of Astronomy \& Astrophysics, 11F of Astronomy-Mathematics Building, AS/NTU, No.1, Sec. 4, Roosevelt Road, Taipei 106319, Taiwan, ROC
}

\date{Received 23 March 2026; accepted 11 August 2026}

  \abstract
  {Complex organic molecules (COMs) are typically observed in the warm inner envelopes of Class 0 protostars (hot corinos). In evolved Class I sources, detecting COMs is challenging, as the emission becomes confined to the compact inner disk and obscured by high dust opacity. However, the FU Orionis-type binary L1551 IRS 5 offers a valuable opportunity to probe disk chemistry, as its outburst luminosity thermally sublimates ices out to large radii.}
  {We sought to spatially resolve the origins of COMs in this system in order to distinguish between central radiative heating and shock-induced chemistry driven by heterogeneous accretion pathways.}
  {We analyzed high-angular-resolution data from the Atacama Large Millimeter/submillimeter Array (ALMA) Science Archive, focusing on COMs (e.g., \ce{CH3OH}, \ce{CH3OCHO}) and the shock tracer \ce{SO}. By comparing their spatial morphologies and kinematics, we diagnosed the excitation conditions associated with different accretion flows.}
  {We find that the COM emission in L1551 IRS 5 arises from both radiative heating and localized accretion shocks. In the northern source, the widespread low-excitation COM emission is primarily consistent with thermal heating by the recent outburst, while  the source also exhibits dual-mode accretion with distinct chemical signatures: COMs along the spiral arm tracing steady equatorial circumbinary disk-to-circumstellar disk accretion and high-excitation COMs with bright SO along the eastern disk surface tracing a vertical envelope-to-disk streamer impact.
  In the southern source, methanol is confined to the southeastern streamer impact site, whereas SO forms a ring-like structure at the tidal truncation radius. We attribute this difference to differential chemical survival timescales.}
  {Our study demonstrates that COM emission in this evolved protostellar binary arises from a complex interplay of thermal heating and shocks driven by multiple accretion and interaction pathways. L1551 IRS 5 thus serves as an ideal laboratory to investigate how shock chemistry and binary-regulated accretion shape the chemical structure of protostellar disks, warranting further high-resolution studies.}

   \maketitle
\nolinenumbers
\setlength{\textfloatsep}{7pt plus 1pt minus 2pt}
\setlength{\dbltextfloatsep}{7pt plus 1pt minus 2pt}
\setlength{\floatsep}{6pt plus 1pt minus 1pt}
\setlength{\dblfloatsep}{6pt plus 1pt minus 1pt}
\setlength{\intextsep}{6pt}
\setlength{\abovecaptionskip}{4pt}
\setlength{\belowcaptionskip}{0pt}

\section{Introduction} \label{sec:intro}

The chemical evolution of protostellar systems is closely linked to their dynamical history. In the earliest Class 0 stage, complex organic molecules (COMs) are frequently detected in the warm inner envelopes, known as hot corinos, where thermal desorption of grain mantles is efficient \citep[e.g.,][]{Ceccarelli:2023aa, Jorgensen:2016aa}. In the more evolved Class~I stage, in contrast, COMs have been detected in far fewer sources. As the disk grows increasingly dense, the warm region where COMs are released into the gas phase becomes confined to the compact inner disk \citep{Kim:2025aa}, and this compact emission suffers from low total flux and high dust opacity \citep[e.g.,][]{De-Simone:2020aa, Yang:2021aa}. Consequently, the chemical inventory of Class I disks and the mechanisms governing their enrichment remain less understood compared to their younger counterparts.

Recent high-sensitivity observations, particularly with ALMA, have begun to reveal that neither the accretion onto the protostar nor the large-scale supply of material to the disk proceeds smoothly. On small scales, accretion onto the protostar and its disk is temporally variable, as evidenced by episodic accretion bursts. On large scales, mass is delivered to the disk anisotropically through asymmetric streamers originating in the surrounding envelope \citep[see review by][]{Pineda:2023aa}.
These dynamic inflows and bursts can significantly alter the thermal structure and chemical abundance profiles of the disk \citep{Lee:2019aa}.
Furthermore, accretion flows can induce shocks at the disk-envelope interface, sputtering dust grains and triggering gas-phase chemistry distinct from thermal sublimation \citep{Sakai:2014aa, Miura:2017aa}.
While shock tracers such as SO and \ce{SO2} have been observed in association with streamers \citep{Garufi:2022aa, Valdivia-Mena:2022aa, Lee:2023vl, Lee:2024aa}, spatially resolving the impact of these shocks on COM emission within the disk remains a major observational challenge.
Specifically, distinguishing shock-induced chemical hotspots from the thermally heated reservoir in the inner disk requires high angular resolution and a target with favorable conditions.

The Class I binary system L1551 IRS 5 is an ideal laboratory to address this question. Located in the Taurus molecular cloud at a distance of 147~pc \citep{Connelley:2008aa}, this system consists of two protostars, Source N and Source S, separated by $\sim$45~au in the north-south direction \citep{Lim:2016tk}. The binary is surrounded by a circumbinary disk (CBD) and drives large-scale molecular outflows \citep{Snell:1980uv} and collimated radio jets \citep{Rodriguez:2003vs, Feeney-Johansson:2023wu, Yang:2022vg}, indicating active ongoing accretion.
Hydrodynamic simulations have shown that the ring-like CBD in the dust-continuum emission consists of two spiral arms caused by the gravitational torque of the binary orbital motion. Inside the Roche lobe, those spiral arms trace channel flow of the materials onto the individual circumstellar disks (CSDs; \citealt{Matsumoto:2019wj}; \citealt{Takakuwa:2020tg}).
As a FU Orionis-type (FUor) object, its elevated luminosity significantly expands the thermal snowline to larger radii ($\sim$10--40 au), rendering sublimated species from ices observable on disk scales that are resolvable with ALMA \citep{Cruz:2025}.

Although recent studies have reported the detection of COMs in this system \citep[e.g.,][]{Bianchi:2020aa, Mercimek:2022vw, Marchand:2024vh, Cruz:2025}, the physical origins of the emission have not been spatially disentangled. Distinguishing whether COMs trace the steady accretion flow, the heated disk surface, or localized shock impacts is observationally difficult. For instance, while \citet{Hsieh:2024aa} recently suggested shock origins for the analogous Class I binary SVS13A using kinematic decomposition, a clear spatial differentiation of these heterogeneous mechanisms has yet to be established due to the angular resolution issue.

In this paper, we present high-angular-resolution observations of COMs and shock tracers toward L1551 IRS 5 using data from the ALMA Science Archive. We spatially resolve the COM emission into distinct components associated with the binary sources, Source N and Source S. By comparing the spatial morphologies and kinematics of various COMs (e.g., methanol, methyl formate) with the shock tracer SO, we aim to disentangle the contributions of thermal heating and accretion shocks. In particular, we report the spatial separation of ``steady equatorial CBD-to-CSD accretion'' through spiral arms and ``vertical envelope-to-disk streamer impacts,'' providing a comprehensive view of how heterogeneous accretion pathways shape the chemical complexity of protostellar disks.

This paper is structured as follows. In Sect.~\ref{sec:data} we describe the archival data and the imaging process. In Sect.~\ref{sec:results} we present the spatial distributions and kinematics of the COMs and SO. In Sect.~\ref{sec:discussion} we discuss the dual-mode accretion scenario in Source N and the chemical dichotomy observed in Source S. Finally, in Sect.~\ref{sec:conclusion} we summarize our conclusions.

\section{Observations and data reduction} \label{sec:data}

We used two Cycle 4 ALMA Science Archive datasets for our analysis: 2016.1.00209.S (PI: Takami) and 2016.1.00138.S (PI: Takakuwa). The details of each observation are fully described in \citet{Kospal:2021aa} and \citet{Takakuwa:2020tg}, respectively. Here, we briefly summarize the observational information relevant to our analysis.

For the project 2016.1.00209.S, L1551 IRS 5 was observed with two extended 12-m configurations, the 7-m array, and total power (TP). Here we used only the 12-m data, observed on 2017 July 24 and 2018 April 5, using the ``tclean'' task in the Common Astronomy Software Applications \citep[CASA;][]{CASA-Team:2022} with Briggs weighting and a robust parameter of 0.5. The synthesized beam after combination is $0\farcs15 \times 0\farcs12$, and the maximum recoverable scale is $8\farcs8$.
We note that the time gap between the two observations is approximately 9 months. Given the known proper motion of L1551 IRS 5 ($\sim$24.5 mas/yr; \citealt{Hernandez-Garnica:2024aa}), the expected positional shift is $\sim$0\farcs017, which is less than 11\% of the synthesized beam size. Therefore, we combined the two datasets without additional astrometric correction, as the shift is significantly smaller than the resolution element.

\begin{figure*}[t]
\centering
\includegraphics[angle=0,width=180mm]{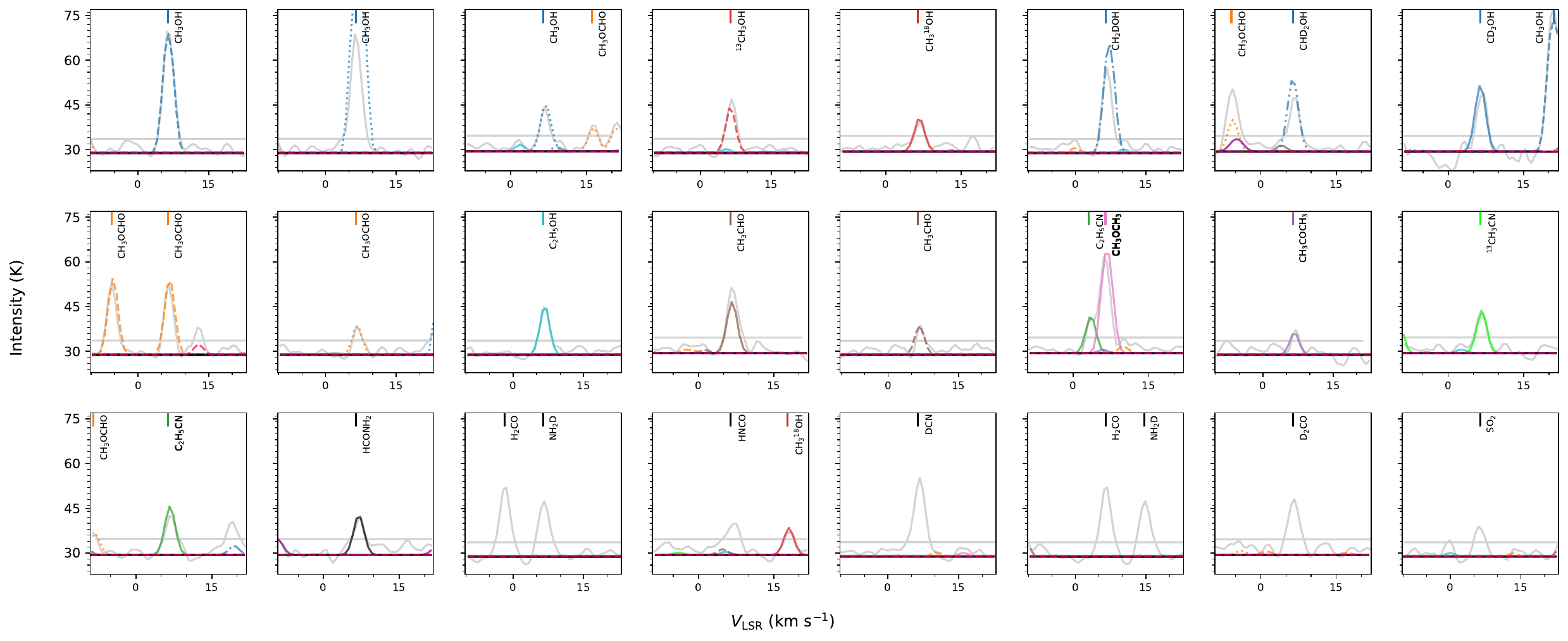}
\caption{Spectra of molecular transitions at (R.A., Decl.) = (4:31:34.162, 18:09:4.91), near position (a) in Figure~\ref{fig:ch3oh}. The LTE model spectra of complex organic molecules ($T_{\rm ex} = 180$~K) are overlaid with different colors and line styles. The order of the molecules is the same as that in Table~\ref{tab:lines}.~\label{fig:spec0}}.
\end{figure*}

We used two spectral windows centered at 216.9 GHz and 232.2 GHz, respectively, each with a bandwidth of 1.8 GHz. The root mean square (rms) noise of the final line cubes is 1.4~mJy~beam$^{-1}$ at the given velocity resolution of 1.3~km~s$^{-1}$. We identified molecular transitions in these windows using the eXtended CASA Line Analysis Software Suite \citep[XCLASS;][]{Moller:2017aa}, which performs radiative transfer modeling under the assumption of local thermodynamic equilibrium (LTE) using the CDMS \citep{Muller:2001uq} and JPL \citep{Pickett:1998aa} catalogs. From the identified transitions, we selected isolated molecular lines that were minimally contaminated by other species (see Figure~\ref{fig:spec0}). The molecular transitions used for the analysis are summarized in Table~\ref{tab:lines}.

The identification of these COMs is consistent with the independent analysis of the same ALMA data by \citet{Cruz:2025}. The few transitions used here that are not in their list have also been securely identified. \ce{CHD2OH}, \ce{CD3OH}, and \ce{HCONH2} were confirmed with a single LTE model that reproduces their multiple uncontaminated transitions, while \ce{NH2D} and \ce{HNCO} were each identified from a single uncontaminated line. A full column density and temperature analysis of the COM inventory will be presented in a forthcoming paper (Kang et al., in prep.).

The observation for project 2016.1.00138.S was conducted on 2017 July 27 in Band 7. The continuum image was constructed with uniform weighting after self-calibration, resulting in a beam size of $0\farcs078 \times 0\farcs065$ and a sensitivity of 0.032 mJy beam$^{-1}$ at 336.1 GHz. We used this continuum image for the analysis during this study. We also utilized the SO ($N_J=8_7 - 7_6$) line transition at 340.7~GHz. The synthesized beam resolution of the SO line cube is $0\farcs11 \times 0\farcs11$, with Briggs weighting and a robust parameter of 0.5. The rms noise of the SO line cube is 2.2 mJy beam$^{-1}$ at a spectral resolution of 0.22 km s$^{-1}$.

\begin{table*}[t]\small
\centering
\caption{Transitions of L1551 IRS 5 identified in this paper.~\label{tab:lines}}
\begin{tabular}{llcc|llcc}
\hline \hline \noalign{\smallskip}
Molecule & Transition & Frequency & $E_{up}$ & Molecule & Transition & Frequency & $E_{up}$ \\
         &            & (MHz)     & (K)      &          &            & (MHz)     & (K)      \\
\hline \noalign{\smallskip}
CH$_3$OH ($v_t=0$)   & $5_{1,4}-4_{2,3}$ E       & 216945.60 & 55.9  & CH$_3$OCH$_3$    & $13_{0,13}-12_{1,12}$    & 231987.86 & 80.9  \\
CH$_3$OH ($v_t=1$)   & $6_{1,5}-7_{2,5}$ A       & 217299.20 & 373.9 & CH$_3$COCH$_3$   & $18_{4,14}-17_{5,13}$ AA & 216000.47 & 111.0 \\
CH$_3$OH ($v_t=1$)   & $9_{6,3}-9_{7,2}$ E       & 232847.15 & 696.5 & $^{13}$CH$_3$CN  & $13_2-12_2$              & 232216.73 & 106.7 \\
$^{13}$CH$_3$OH      & $10_{2,8}-9_{3,7}$ A      & 217399.55 & 162.4 & C$_2$H$_5$CN     & $26_{7,20}-25_{7,19}$    & 233069.31 & 205.4 \\
CH$_3$O$^{18}$H      & $5_{0,5}-4_{0,4}$ E       & 231686.68 & 46.2  & \ce{HCONH2}        & $11_{2,10}-10_{2,9}$     & 232273.65 & 79.0  \\
CH$_2$DOH            & $12_{0,12}-11_{1,11}$ $e_0$ & 216129.63 & 166.5 & NH$_2$D          & $3_{2,2}-3_{1,2}$        & 216562.72 & 119.6 \\
CHD$_2$OH            & $13_{1}-12_{2}$           & 231714.69 & 186.8 & HNCO             & $28_{1,28}-29_{0,29}$      & 231873.26 & 469.9  \\
CD$_3$OH             & $6_{2,4}-6_{-1,6}$        & 231292.38 & 62.0  & DCN              & $J=3-2$                  & 217238.54 & 20.9  \\
CH$_3$OCHO ($v=0$)   & $18_{2,16}-17_{2,15}$ E   & 216830.20 & 105.7 & H$_2$CO          & $9_{1,8}-9_{1,9}$        & 216568.65 & 174.0 \\
CH$_3$OCHO ($v_t=1$) & $17_{4,13}-16_{4,12}$ A   & 217312.63 & 290.0 & D$_2$CO          & $4_{0,4}-3_{0,3}$        & 231410.23 & 27.9  \\
C$_2$H$_5$OH         & $13_{0,13}-12_{0,12}$     & 216415.67 & 130.6 & SO$_2$           & $22_{2,20}-22_{1,21}$    & 216643.30 & 248.4 \\
CH$_3$CHO ($v=0$)    & $12_{4,9}-11_{4,8}$ E     & 231506.29 & 108.3 & SO               & $N_J=8_7-7_6$            & 340714.16 & 81.2  \\
CH$_3$CHO ($v_t=1$)  & $11_{1,10}-10_{1,9}$ E    & 216294.83 & 269.6 &                  &                          &           &       \\
\hline
\end{tabular}
\begin{minipage}{0.9\textwidth}
\vspace{1ex}
\footnotesize
Note. Transitions are designated as $J_{K_a,K_c} - J'_{K'_a,K'_c}$ unless otherwise indicated.
\end{minipage}
\end{table*}

\section{Results} \label{sec:results}
\subsection{Spatial distributions and chemical structures} \label{sec:spatial}

\subsubsection{Overview of molecular emission}
\begin{figure*}[t]
\centering
\includegraphics[angle=0,width=170mm]{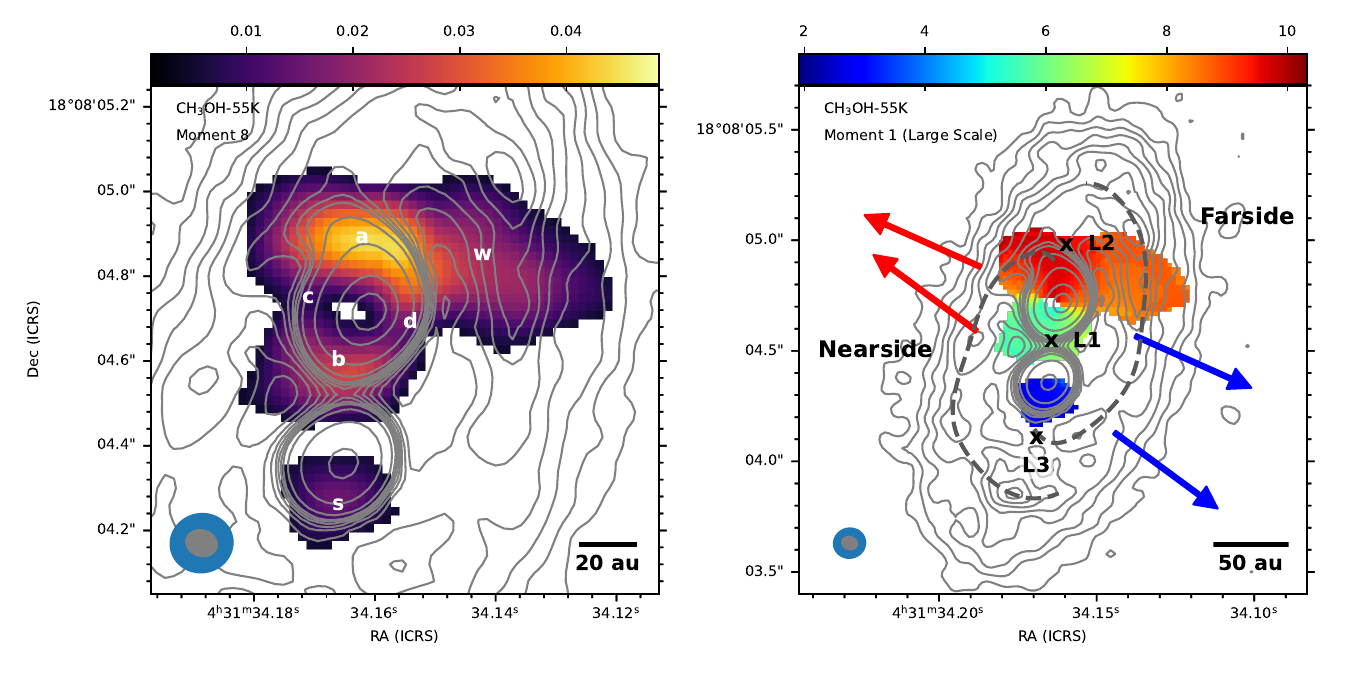}
\caption{Peak intensity (left) and intensity-weighted velocity (right) maps of the CH$_{3}$OH (216.9~GHz, $E_\mathrm{up} = 55$~K) emission overlaid with 0.9~mm continuum contours.
The continuum contour levels are 3, 6, 9, 12, 15, 18, 21, 24, 27, 30, 40, 80, 200, and 300$\sigma$, where $\sigma = 0.31$~mJy~beam$^{-1}$ in both panels.
The synthesized beams of the CH$_{3}$OH and continuum emissions are indicated by blue and gray ellipses, respectively, in the bottom-left corners. Left panel: Moment 8 map (color scale in units of Jy~beam$^{-1}$~K~km~s$^{-1}$). The labels mark the regions where the spectra discussed in Sect.~\ref{sec:spectra} were extracted: ``a,'' ``b,'' ``c,'' and ``d'' represent the north, south, east, and west regions of the northern CSD, respectively. Further, ``w'' denotes the CH$_{3}$OH western extension, and ``s'' indicates the CH$_{3}$OH spot associated with Source S. Right panel: Zoomed-out moment 1 map displaying large-scale kinematic and structural features (color scale in units of km~s$^{-1}$). Receding and approaching jets from the binary are indicated by red and blue arrows, respectively. Text labels denote the nearside and farside of the CBD and CSDs, inferred from the jet inclinations. The dark gray dashed lines outline the approximate continuum spiral arms estimated from simulations by \citet{Takakuwa:2020tg}. Lagrangian points (L1, L2, and L3) are marked with crosses, assuming equal stellar masses; L1 is midway between the two sources (N and S), and L2 and L3 are located at $\sim$1.2$a$ from L1, where $a$ is the binary separation. A scale bar is shown in the bottom-right of each panel. \label{fig:ch3oh}}
\end{figure*}

Figure~\ref{fig:ch3oh} illustrates the spatial and velocity distributions of the $\text{CH}_3\text{OH}$ transition at 216.9~GHz ($E_{\text{up}} = 55$~K). The methanol emission is primarily concentrated around Source~N. It is brightest in the north of Source N and extends toward the CBD to the east and west. At the center, the emission is significantly obscured due to the high optical depth of the dust continuum. Methanol is also detected near Source~S, although it is offset toward the south.

The velocity map of Source~N exhibits a gradient consistent with the rotation of the CSD. The emission from the northern source is present at $+4.0 < v_{\rm LSR} < +11.85$~km~s$^{-1}$, and the emission from the southern source occurs at $+1.7 < v_{\rm LSR} < +3.7$~km~s$^{-1}$. The directions of the blue- and red-shifted ionized jets \citep{Feeney-Johansson:2023wu} are indicated by blue and red arrows, respectively. These jet orientations suggest that the protostellar disks are inclined, with the eastern and western sides representing the nearside and farside of the disk, respectively. The continuum spirals proposed by \citet{Takakuwa:2020tg} are indicated by dark gray dashed lines, along with the Lagrangian points calculated assuming equal stellar masses.

Various molecular species were identified toward Source N (Figure~\ref{fig:mom0}), comprising six O-bearing complex organic molecules (COMs) including methanol, three N-bearing COMs, and seven simple molecules. Specifically, we detected methanol (CH$_3$OH) along with its isotopologs ($^{13}$CH$_3$OH, CH$_3^{18}$OH) and deuterated species (CH$_2$DOH, CHD$_2$OH, and CD$_3$OH), methyl formate (CH$_3$OCHO), ethanol (C$_2$H$_5$OH), acetaldehyde (CH$_3$CHO), dimethyl ether (CH$_3$OCH$_3$), and acetone (CH$_3$COCH$_3$). The N-bearing species and simple molecules include methyl cyanide ($^{13}$CH$_3$CN), ethyl cyanide (C$_2$H$_5$CN), formamide (\ce{HCONH2}), NH$_2$D, HNCO, DCN, H$_2$CO, D$_2$CO, SO$_2$, and SO. The western extension of the molecular emission from Source N, which is clearly seen in methanol, is also prominent in the DCN and H$_2$CO maps.

Toward Source S, we identified CH$_3$OH, CH$_3$OCHO, CH$_3$OCH$_3$, CH$_3$COCH$_3$, C$_2$H$_5$CN, NH$_2$D, DCN, H$_2$CO, and SO. Notably, the morphologies of dimethyl ether and SO differ from those of other molecules; their emission covers the entire area of Source S and exhibits a more spatially extended distribution across the circumbinary system. Further details regarding the SO emission and its kinematics are described in Sect.~\ref{sec:streamer_connection}

\subsubsection{Source N: Structural asymmetries}

\begin{figure*}[t]
\centering
\includegraphics[angle=0,width=180mm]{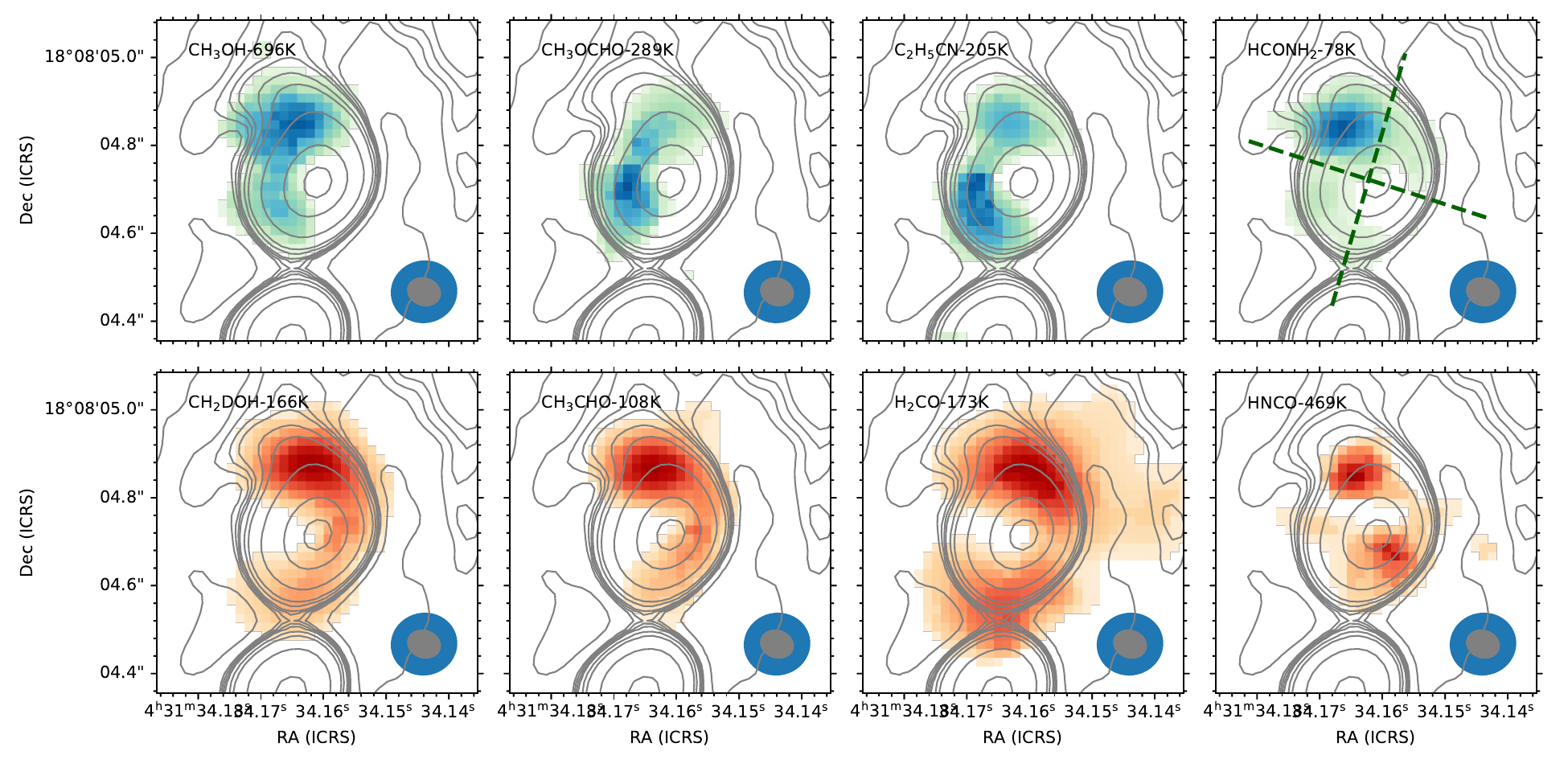}
\caption{
Zoomed-in moment 0 maps centered at Source N (RA, Dec = 4:31:34.161, 18:08:04.72). Top panels: Molecules brighter on the nearside of the disk. Bottom panels: Molecules brighter on the farside of the disk. The gray contours represent the dust continuum at levels of 4, 12, 20, 22, 24, 26, 32, 40, 80, 200, and 320$\sigma$, where $\sigma = 0.31$~mJy~beam$^{-1}$. The dashed lines in the top-right panel indicate the lines along which the radial intensity profiles in Figure~\ref{fig:radial} and Figure~\ref{fig:radial2} are retrieved.
In the bottom-right of each panel, the beam size of the molecular lines is shown in blue, and the 0.9~mm dust continuum beam is shown in gray.
~\label{fig:anticorr}}
\end{figure*}

A notable feature in Figure~\ref{fig:mom0} is the asymmetric molecular distribution between the nearside and farside of the disk in Source~N. To highlight this trend, Figure~\ref{fig:anticorr} compares two groups of molecules exhibiting distinct spatial patterns.

The molecules shown in the bottom panels of Figure~\ref{fig:anticorr} (\ce{CH2DOH}, \ce{CH3CHO}, \ce{H2CO}, and \ce{HNCO}) are predominantly bright in the north and distributed along the western rim of Source~N, which corresponds to the farside of the disk. In contrast, these species are either undetected or significantly fainter along the eastern (nearside) rim. This distribution is consistent with a scenario where molecules are released from dust grains into the gas phase following an FUor-type outburst in Source~N, inclined by $\sim$40\degr \citep{Cruz-Saenz-de-Miera:2019aa}. In an inclined and flared disk geometry, emission originating from the nearside of the midplane can be obscured by the cold, optically thick outer layers of the disk or envelope.

Conversely, the molecules presented in the top panels (\ce{CH3OH}, \ce{CH3OCHO}, \ce{C2H5CN}, and \ce{HCONH2}) exhibit the opposite spatial trend, being primarily detected along the eastern (nearside) rim while remaining faint or undetected toward the west. It is noteworthy that these specific transitions are characterized by significantly high excitation requirements ($E_{\text{up}} > 200$~K, with the exception of \ce{HCONH2}). Interestingly, the lower $E_{\text{up}}$ transitions of the same species---such as \ce{CH3OH} ($E_{\text{up}} = 55$~K) and \ce{CH3OCHO} ($E_{\text{up}} = 105$~K) shown in Figure~\ref{fig:mom0}---tend to follow the morphology of the molecules in the bottom panels, which are prominent along the western (farside) rim.

Another notable feature observed in Figures~\ref{fig:mom0} and \ref{fig:anticorr} is that the brightest emission in the northern region is offset eastward relative to the major axis of Source~N (indicated by the dotted line in the top-right panel of Figure~\ref{fig:anticorr}). This location coincides with the region where one of the two CBD spiral arms predicted by the hydrodynamic simulations of \citet{Takakuwa:2020tg} connects to Source~N. This continuum feature is highlighted by the dust continuum contours overlaid in Figures~3 and 4. The underlying causes of the asymmetric emission between the nearside and farside as well as the enhanced molecular emission at the junction of the spiral arm are addressed further in the discussion section (Sect.~\ref{sec:discussion}).

Figure~\ref{fig:radial} presents the radial intensity profiles of representative molecular transitions along the major and minor axes of Source~N. The bottom panels show the intensity distributions along the major axis, where most molecules exhibit peaks at the disk rims. For \ce{CH3OH} ($E_{\text{up}} = 55$~K) and \ce{CH2DOH}, the emission peaks at a radius of $\sim$22~au at a distance of 147~pc, with the northern peak being notably brighter than the southern one. In contrast, species with lower column densities, such as $^{13}\text{CH}_3\text{OH}$, $^{13}\text{CH}_3\text{CN}$, and \ce{C2H5CN}, show relatively symmetric intensities between the two rims.

Most molecular emission is confined within 40~au, which is consistent with the dust continuum boundary of CSD. This suggests that the entire protostellar material of Source~N, truncated by tidal forces, has reached temperatures above $\sim$100~K, sufficient to sublimate COMs from dust grains. Indeed, the peak 0.9-mm brightness temperature of the CSD in Source N is known to exceed 260 K (Takakuwa et al. 2020). Interestingly, the SO emission peaks near the Lagrangian point between Sources~N and S, suggesting the presence of shocks potentially driven by material exchange between the binary components.

The top panels of Figure~\ref{fig:radial} show the deprojected radial profiles along the minor axis, assuming an inclination of $40^\circ$. While the \ce{CH3OH} (55~K) line exhibits peaks at both the nearside and farside rims ($\sim$22~au), other species show distinct asymmetries; for instance, \ce{CH2DOH} and $^{13}\text{CH}_3\text{CN}$ are detected only on the farside, whereas \ce{C2H5CN} and $^{13}\text{CH}_3\text{OH}$ are seen only on the nearside, consistent with the trends noted in the previous section. To further emphasize this, Figure~\ref{fig:radial2} compares the relative intensities between the nearside and farside rims by separating molecules that are brighter on the nearside (top) from those brighter on the farside (bottom). It is evident that most molecules in the bottom panels have $E_{\text{up}} < 170$~K (with the exception of simple molecules, such as \ce{SO2} and \ce{HNCO}), while those in the top panels generally possess $E_{\text{up}} > 160$~K (except for \ce{HCONH2}).

\subsection{Kinematics}\label{sec:kinematics}

\subsubsection{Keplerian model analysis}
To investigate kinematic deviations from pure disk rotation in Source N, we constructed a velocity residual map (Figure~\ref{fig:v_residual}) using the Moment 1 map of the \ce{CH3OH} ($E_{\text{up}} = 373$~K) transition, which is expected to be more optically thin than the 55~K line. The residuals were obtained by subtracting a Keplerian rotation model derived using stellar parameters from previous studies: a stellar mass ($M_*$) of $0.25~M_\odot$, a position angle (PA) of $156^\circ$, and an inclination of $40^\circ$ \citep{Cruz-Saenz-de-Miera:2019aa, Takakuwa:2020tg}. The systemic velocity ($V_{\text{sys}}$) was optimized to $+7.4$~km~s$^{-1}$ by minimizing the residuals within the radial range of $18 < r < 40$~au. We note that this value is slightly higher than the $+6.85$~km~s$^{-1}$ reported by \citet{Takakuwa:2020tg}. The central region within $r < 18$~au was excluded from the analysis, as the velocity information in this innermost region is likely distorted by significant dust attenuation.

As shown in Figure~\ref{fig:v_residual}, the majority of the emission region exhibits small residuals within $\pm 0.5$~km~s$^{-1}$, suggesting that the bulk motion is well described by Keplerian rotation. However, distinct non-Keplerian excesses are identified in specific regions. A significant positive velocity excess is observed in the northeastern region ($0\arcsec < \Delta\text{RA} < +0\farcs2$, $+0\farcs1 < \Delta\text{Dec} < 0\farcs3$), where the CBD spiral arm connects to the protostellar disk of Source N. In this region, the residual velocity gradually increases from the inner disk toward the outer spiral arm, reaching values $> +2.0$~km~s$^{-1}$.

In addition, the western molecular extension displays a positive excess, while the eastern disk region (offset $\approx +0\farcs15, 0\arcsec$) exhibits a prominent negative excess. These features indicate the presence of significant non-Keplerian velocity components locally associated with the shock interactions, together with the multiple velocity components that are shown in Sect.~\ref{sec:spectra}.

\subsubsection{Spectral analysis}~\label{sec:spectra}

Figure~\ref{fig:spec} compares the molecular spectra of methanol lines with low and high excitation energies ($E_{\rm up} = 55$ and $373$~K), \ce{HCONH2}, \ce{DCN}, and \ce{SO} toward the multiple regions of the binary system marked in Figure~\ref{fig:ch3oh} (left). In general, the spectral profiles of different molecular species exhibit similarities at each position and are relatively well characterized by a single velocity component, with the notable exception of the nearside profile (region c). This similarity in line profiles suggests a common spatial origin for these molecular emissions.

In the case of the nearside spectra (region c), however, multiple velocity peaks are detected at $+6$~km~s$^{-1}$, $\sim +8$~km~s$^{-1}$, and possibly around $+10.5$~km~s$^{-1}$. Interestingly, while the low-excitation \ce{CH3OH} (55~K) emission peaks at the $+8$~km~s$^{-1}$ component, close to the expected rotation velocity, the high-excitation \ce{CH3OH} (373~K) line peaks at $+6$~km~s$^{-1}$, which likely contributes to the blue-shifted excess observed in the residual map (Figure~\ref{fig:v_residual}). The dominance of the $+6$~km~s$^{-1}$ component in the high-excitation line is distinct from other regions where both transitions typically show comparable intensities or stronger \ce{CH3OH} (55~K) emission.

The region where the 373~K line exceeds the 55~K line is outlined by the green boundary in the residual panel of Figure~\ref{fig:v_residual}. It spreads over a considerably wide area covering much of the eastern disk surface, including but not limited to the blue-shifted non-Keplerian region. The two velocity components are also spatially distinct in the channel maps of individual COM transitions such as \ce{CH3OCHO} (290~K) and \ce{C2H5CN} (Appendix~\ref{app:chan}). The blue-shifted component appears at $V_{\rm LSR}\simeq+6$~km~s$^{-1}$ on the eastern disk surface, and the red-shifted component associated with the spiral-arm root appears at $V_{\rm LSR}\simeq+8$ to $+10$~km~s$^{-1}$.

We interpret this reversal in the relative intensities of the high-excitation and low-excitation \ce{CH3OH} lines between the two components as evidence that they have different excitation temperatures, with the $+6$~km~s$^{-1}$ component being warmer. A quantitative temperature estimate is difficult with the three methanol transitions presented here, because the 55~K line is likely optically thick and the 373 and 696~K lines alone cannot reliably constrain the rotation temperature. We therefore limit ourselves to the qualitative interpretation that the nearside is warmer, supported by the fact that the highest-excitation methanol line ($E_{\rm up}=696$~K) is detected only on the nearside rim (Figure~\ref{fig:radial2}, top) and that the 55/373~K intensity reversal likewise occurs only there (Figure~\ref{fig:v_residual}).

For \ce{HCONH2} and \ce{DCN}, the emission near the systemic rotation velocity is relatively weak; instead, these lines show double peaks at $+6$ and $+10.5$~km~s$^{-1}$. Since these molecules are often considered to trace shocks, or to be enhanced in shocked regions \citep[e.g.,][]{Mendoza:2014aa, Lopez-Sepulcre:2019vh, Busquet:2017aa}, this spectral morphology can be interpreted as the presence of both approaching and receding shocked components in this region. The positive velocity component is likely associated with the positive excess originating from the spiral arm inflow seen in Figure~\ref{fig:v_residual}.

Another notable observational feature is the intensity ratio between \ce{SO} and methanol. At the northern and southern rims of Source N (a and b), the intensities of \ce{SO} and methanol are comparable. In contrast, \ce{SO} is approximately 2--5 times brighter than methanol in other regions. This trend suggests potential shock-related activities in the western extension (w), Source S (s), and the western disk region (d), which are discussed in detail in the following section on \ce{SO} emission.

\begin{figure}[t]
\centering
\includegraphics[width=80mm]{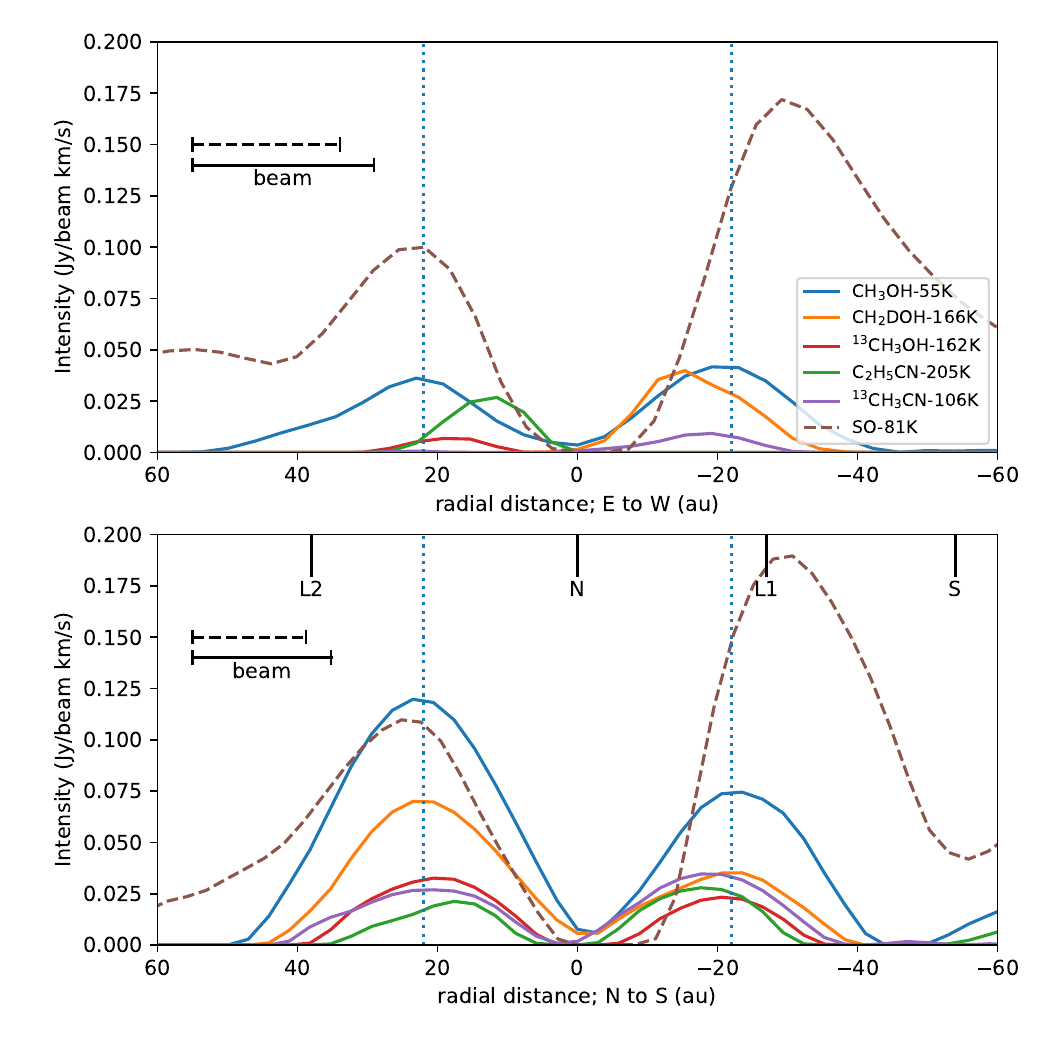}
\caption{Radial intensity profiles of molecules along the minor (top) and major (bottom) axes of the northern CSD when adopting a distance of 147~pc and an inclination of 40$\degr$. Vertical dotted lines mark the CH$_3$OH (55~K) peak at 22~au. COMs are shown as solid lines and the SO as a dashed line; colors are indicated in the legend. Deprojected beam sizes are shown in the upper-left corners. Lagrangian points and the positions of Sources~N and~S are marked on the top axis of the bottom panel (see Fig.~\ref{fig:ch3oh} for definitions).
~\label{fig:radial}}
\end{figure}

\begin{figure}[t]
\centering
\includegraphics[width=80mm]{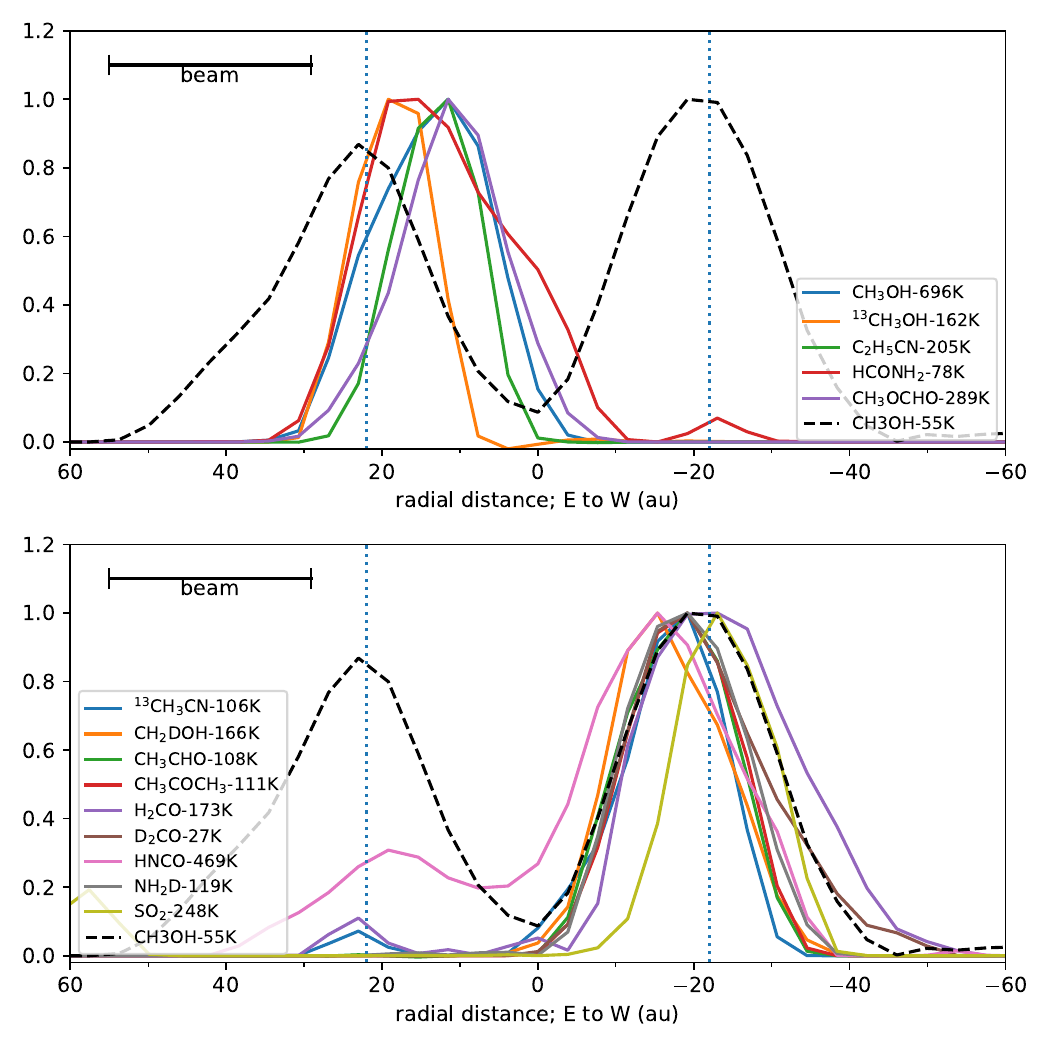}
\caption{Same as the top panel of Figure~\ref{fig:radial}, but for different molecules tracing the nearside and farside of the disks in the top and bottom panels, respectively. CH$_3$OH-55K is presented using a dashed black line for reference.
~\label{fig:radial2}}
\end{figure}

\begin{figure*}[ht!]
\centering
\includegraphics[width=\textwidth]{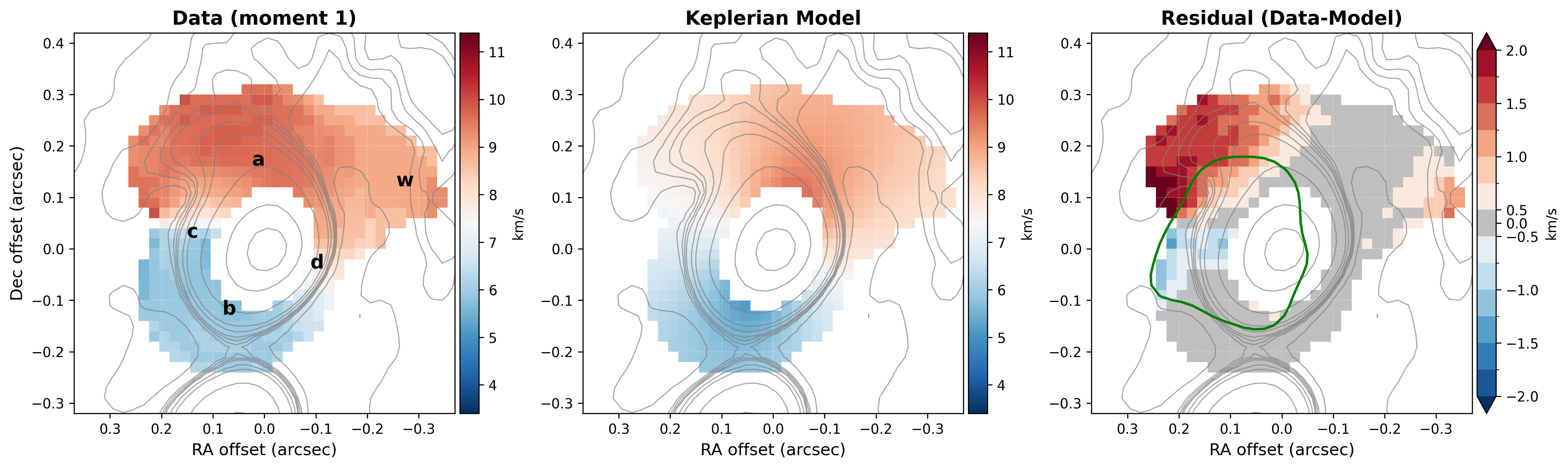}
\caption{Comparison between the observed and modeled velocity fields of \ce{CH3OH} ($E_{\text{up}} = 373$~K). From left to right, the Moment 1 map, the best-fit Keplerian model, and the residual map. The central 18~au region is masked. In the residual map (right), the gray area indicates residuals $|v_{\text{data}} - v_{\text{model}}| < 0.5$~km~s$^{-1}$, highlighting the non-Keplerian excess in the northern region. Contours are the same as in Fig.~\ref{fig:mom0}. The green line in the right panel outlines the smoothed boundary of the region where the $E_{\rm up}=373$~K \ce{CH3OH} line exceeds the 55~K line at $V_{\rm LSR}=5$--$7$~km~s$^{-1}$.
~\label{fig:v_residual}}
\end{figure*}

\begin{figure*}[t]
\centering
\includegraphics[width=0.95\textwidth]{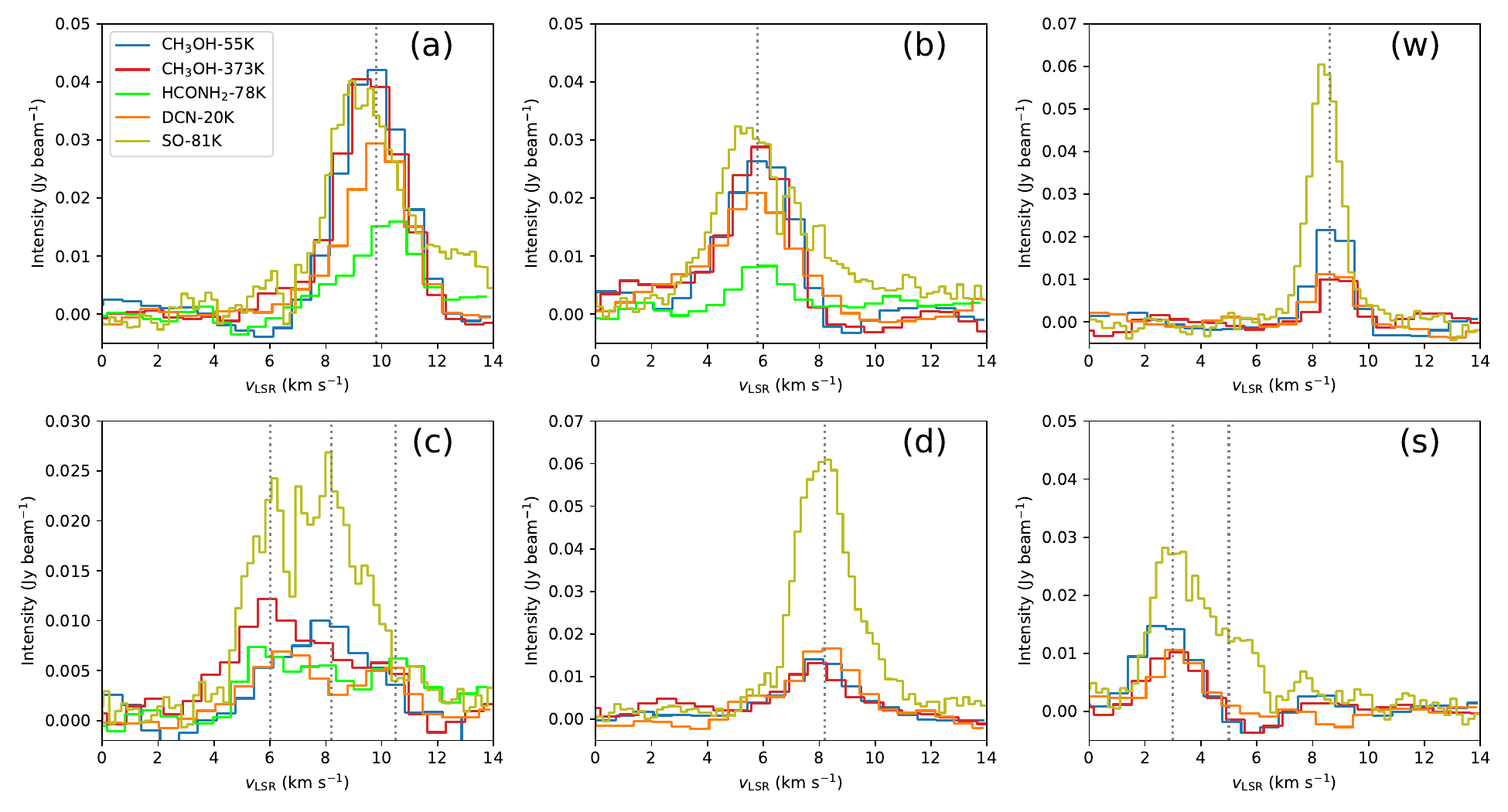}
\caption{
Spectral profiles of \ce{CH3OH} ($E_{\text{up}} = 55$ and $373$~K), \ce{HCONH2} ($E_{\text{up}} = 78$~K), \ce{DCN} ($E_{\text{up}} = 20$~K), and \ce{SO} ($E_{\text{up}} = 81$~K). The spectra were extracted from the northern (a), southern (b), eastern (c), and western (d) rims of Source N and the western extension (w) and the COMs peak of Source S (s), as indicated in Figure~\ref{fig:ch3oh}. Vertical dotted lines in each panel indicate representative velocities. The color legend for each molecule is provided in panel (a).
}
\label{fig:spec}
\end{figure*}

\subsection{Large-scale streamers and their connection to COMs} \label{sec:streamer_connection}
\begin{figure*}[t]
\centering
\includegraphics[width=180mm]{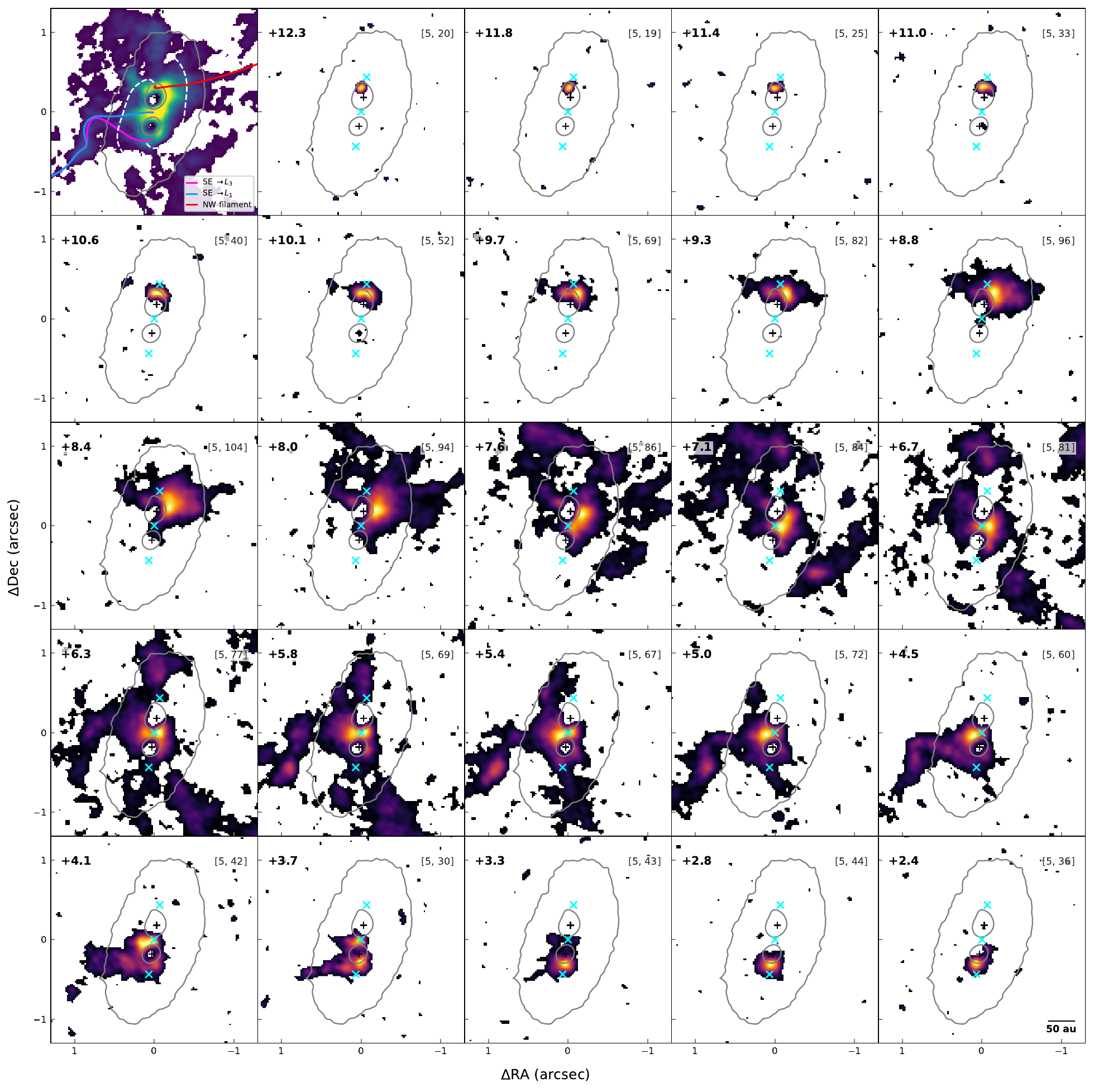}
\caption{Zoomed-in \ce{SO} channel maps of the central region at a velocity sampling of $0.43$~km~s$^{-1}$. The first panel shows the \ce{SO} moment~0 with the three streamer paths used for the position--velocity diagrams (Figure~\ref{fig:pv_se}), SE\,$\to L_3$ (magenta), SE\,$\to L_1$ (blue), and the NW filament (red), and the two continuum spiral arms (white dashed; \citealt{Takakuwa:2020tg}). Each channel panel uses an independent linear color scale, with the brightness-temperature range (K) in brackets and the LSR velocity (km~s$^{-1}$) in the upper left. Black and cyan crosses mark the positions of the protostars and the Lagrangian points ($L_2$, $L_1$, $L_3$ from top to bottom); gray contours show the 0.9~mm continuum at 6 and $50\sigma_{\rm cont}$ ($\sigma_{\rm cont}=0.31$~mJy~beam$^{-1}$); a 50~au scale bar is shown in the last panel.\label{fig:chan_so_in}}
\end{figure*}

The large-scale \ce{SO} emission around the binary is presented in the channel maps of Figure~\ref{fig:chan_so_in} and Appendix~\ref{app:chan}. We identify several curved, filamentary structures extending up to $\sim 600$~au. The overall velocity gradient of the \ce{SO} emission, blue-shifted toward the southeast and red-shifted toward the northwest relative to the systemic velocity ($V_{\text{sys}} = +6.4$~km~s$^{-1}$), generally corresponds to the rotation of the CBD. On smaller scales (Figure~\ref{fig:chan_so_in}, first panel), \ce{SO} shows ring-like structures around both sources, with the brightest emission found along the western rim of Source N and near the $L_1$ region (the transition zone between the two protostars).

The southeastern \ce{SO} feature, which appears from the blue-shifted velocity of $+6.3$~km~s$^{-1}$, bifurcates as it approaches the binary. One branch connects to the south of Source~S near the Lagrangian point $L_3$ at $V_{\rm LSR}\simeq+4.5$ to $+3$~km~s$^{-1}$, and the other connects to the $L_1$ region at $V_{\rm LSR}\simeq+6$ to $+5$~km~s$^{-1}$ (Figures~\ref{fig:chan_so_in} and \ref{fig:chan_so_out}). In the channel maps the southeastern emission shifts position continuously from channel to channel, tracing velocity-coherent strands rather than a single static, broad component. The broad \ce{SO} profile at a single position (e.g., spanning $2$--$6$~km~s$^{-1}$ at position~(s) in Figure~\ref{fig:spec}) reflects the line-of-sight superposition of these strands together with post-shock gas at the $L_3$ landing site. Since the Lagrangian points correspond to effective potential minima, as simulated by \citet{Takakuwa:2020tg}, it is natural for the infalling material to flow through these points. The $L_3$ branch spatially connects to the COMs-rich region of Source~S, and the $L_1$ branch is adjacent to the blue-shifted COMs region of Source~N (Appendix~\ref{app:chan}).

\begin{figure*}[t]
\centering
\includegraphics[width=175mm]{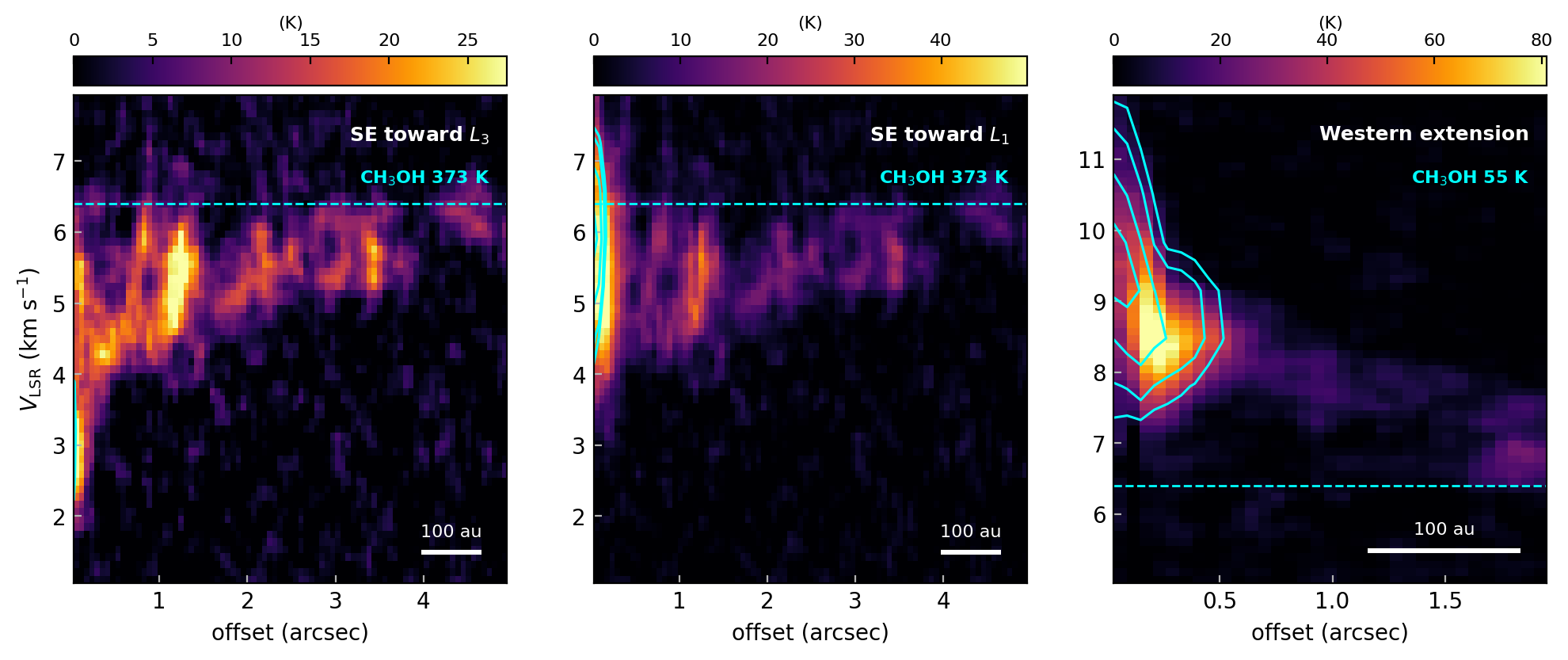}
\caption{Position--velocity diagrams of \ce{SO} along the three streamer paths, extracted with a narrow slit of width $0.2\arcsec$ following the streamer paths drawn in the first panel of Figure~\ref{fig:chan_so_in}. The offset is measured along each path from the source-side end ($0\arcsec$), and the cyan dashed line marks the systemic velocity $V_{\rm sys}=+6.4$~km~s$^{-1}$. (Left) Southeastern branch toward $L_3$, which approaches the source velocity near $+3$~km~s$^{-1}$. (Middle) Southeastern branch toward $L_1$, which approaches the source velocity near $+5$ to $+6$~km~s$^{-1}$. Both southeastern branches show emission that is velocity-coherent along the slit and that connects to the binary at distinct velocities, supporting the interpretation that the southeastern feature is a bifurcating strand rather than diffuse, broad shocked gas. (Right) Northwestern (western) filament, which is entirely red-shifted, reaching $\sim+8.5$~km~s$^{-1}$ at the western rim of Source~N and converging toward the systemic velocity at larger offsets. Cyan contours show the PV of a representative COM transition extracted along the same path, \ce{CH3OH} (373~K) for the two southeastern panels and \ce{CH3OH} (55~K) for the western panel, at 15, 30, 60, and 90\% of its peak. In each panel the COM emission appears at the small-offset end of the path, at the velocity where the \ce{SO} strand arrives. A 100~au scale bar is shown in the bottom-right of each panel.\label{fig:pv_se}}
\end{figure*}

The position--velocity (PV) diagrams extracted along the two branches (Figure~\ref{fig:pv_se}, first and second panels) show that each branch is velocity coherent along its path and that the velocity gradient steepens toward the sources, reaching $\simeq+3$~km~s$^{-1}$ at the $L_3$ landing site and $\simeq+5$ to $+6$~km~s$^{-1}$ at the $L_1$ region. The \ce{CH3OH} contours overlaid on the PV diagrams show that the COM emission appears at the end of each path, at the velocity where the \ce{SO} strand arrives. The nature of these velocity gradients is discussed in Sect.~\ref{sec:shocks}.

On the other hand, an extended \ce{SO} filament, the western extension of Source~N, is seen in the $+8.0$ to $+9.3$~km~s$^{-1}$ channels (Figures~\ref{fig:chan_so_in} and \ref{fig:chan_so_out}). It coincides in position with the brightest \ce{SO} peak in the system ($T_{\rm b} \approx 90$--$100$~K) and appears morphologically associated with the western extensions of methanol (Figure~\ref{fig:ch3oh}) and DCN (Figure~\ref{fig:mom0}), although they do not appear to be kinematically connected to the large-scale red-shifted \ce{SO} farther to the west. We note that the \ce{SO} peak is shifted to the west compared to the \ce{CH3OH} (55~K) emission, as shown in Figure~\ref{fig:radial} (top). These features are also seen in the PV diagram along this filament (Figure~\ref{fig:pv_se}, right panel). The brightest \ce{SO} peak appears at offset $\simeq0\farcs2$ and $V_{\rm LSR}\simeq+8.5$~km~s$^{-1}$, offset from the COM peak, and the velocity declines smoothly toward the systemic velocity with distance from the rim. The most western arc-like \ce{SO} feature (offset $\simeq2\arcsec$, $V_{\rm LSR}\simeq+6.8$~km~s$^{-1}$) appears as a velocity component distinct from this coherent structure. Given the lack of significant methanol or COMs enhancement at the exact \ce{SO} peak position, this specific \ce{SO} flow does not appear to be the main cause of the observed COMs emission in the disk (Sect.~\ref{sec:shocks}).

Finally, the \ce{SO} emission in the higher velocity range of $+9.8$ to $+12$~km~s$^{-1}$ is found primarily to the north of Source~N, around the root of the northern continuum spiral arm (marked by the white dashed curves in the first panel of Figure~\ref{fig:chan_so_in}). In this velocity range, and also at $V_{\rm LSR}\simeq+8$ to $+10$~km~s$^{-1}$, the \ce{SO} emission at and around the arm root is considerably fainter than that at the streamer-related regions, such as the $L_1$ region and the western rim of Source~N (see the \ce{SO} moment~0 map in the first panel of Figure~\ref{fig:chan_so_in} and the \ce{SO} panel of Fig.~\ref{fig:mom0}). The origin of this contrast is discussed in Sect.~\ref{sec:shockintensity}.

\section{Discussion} \label{sec:discussion}

\subsection{Origin of COMs in L1551 IRS 5}

The resolved ALMA images of the close binary system L1551 IRS 5 suggest that multiple mechanisms---thermal heating from central source outbursts, shocks from mass accretion via the CBD, or direct mass accretion to the CSD via streamers from the envelope---may contribute to the emission of COMs. We examine each of these mechanisms in turn below.

\subsubsection{Thermal heating from central sources}

We first investigated whether the sublimation of ice mantles by radiative heating from the central protostars can explain the observed COMs distribution.
The radius of the water snowline ($R_{\text{snow}}$), where the dust temperature reaches the sublimation temperature of water ice ($T_{\text{sub}} \sim 100$~K), can be estimated using the standard luminosity scaling relation for radiatively heated envelopes \citep[e.g.,][]{Bisschop:2007tm, vant-Hoff:2022tz}: $R_{\text{snow}} \approx R_0 (L_{\text{bol}}/1~L_\odot)^{0.5}$, where we adopt a scaling factor of $R_0 = 15.4$~au, the value corresponding to $T_{\rm sub}=100$~K in \citet{Bisschop:2007tm}.
Adopting the bolometric luminosities of $L \approx 30~L_\odot$ for Source N and $10~L_\odot$ for Source S \citep{Liseau:2005ut}, the theoretical water snowline radii are calculated to be approximately 84~au and 49~au, respectively.

These radii are upper limits.
The presence of a dense disk can significantly shield the midplane from central irradiation, resulting in a smaller effective snowline radius compared to the optically thin envelope approximation.
Recently, \citet{Kim:2025aa} demonstrated using radiative transfer modeling that the dense disk structure in an envelope-plus-disk system effectively blocks central radiation, reducing the snowline scaling factor to approximately one-third of that in an envelope-only model. Consistent with this, \citet{Murillo:2022aa} also showed that the structure of the protostellar core and disk strongly affects the predicted snowline locations.
Adopting this result, the expected water snowline radii for Source N and Source S decrease to roughly $\sim 28$~au and $\sim 16$~au, respectively.

In Source~N, the bright emission of the low-$E_{\text{up}}$ COMs and simple molecules (e.g., \ce{H2CO}, \ce{D2CO}, \ce{HNCO}, \ce{NH2D}) can be explained by the expansion of the water snowline driven by the recent accretion outburst. The disk-shielded snowline of $\sim 28$~au is comparable to the tidal truncation radius of the CSD \citep{Artymowicz:1994aa, Harris:2012aa}, which is also recognizable in the dust continuum contours, so the outburst heating can thermally sublimate COMs across the truncated disk. The typical COMs peak at $\sim 22$~au (Figure~\ref{fig:radial}) rather than at the center because the dense inner disk is optically thick, and the emission becomes visible only where the dust turns optically thin. The truncated CSDs of the two sources almost touch each other, so material belonging to Source~S lies just outside the truncation radius of the Source~N CSD. This material shows no trace of COM emission that would indicate heating by Source~N. This suggests that the COM boundary of Source~N is set by its snowline, determined by the luminosity of the central source, rather than by the truncation of the disk material. This in turn indicates that the extended emission of the more abundant molecules, in particular the western extension where some molecules reach $\sim 40$--$70$~au (e.g., \ce{CH3OH} at 55~K, \ce{H2CO}, and \ce{DCN}, Figures~\ref{fig:ch3oh} and \ref{fig:mom0}), lies beyond the snowline and cannot be explained by radiative heating from the central source alone. This instead points to an additional localized mechanism such as accretion shocks (Sect.~4.1.2).

For Source S, although the theoretical disk-plus-envelope snowline radius ($\sim 16$~au) suggests that COMs should be sublimated within the inner protostellar disk, no significant COMs emission is detected around the central position (except for the localized southern enhancement).
This absence implies that the actual COMs emitting region in Source S is  obscured by the high dust optical depth of the dense disk \citep[e.g.,][]{De-Simone:2020aa}.

\subsubsection{Shocks due to heterogeneous accretion pathways}\label{sec:shocks}

\begin{figure*}
    \centering
    \includegraphics[width=0.85\textwidth]{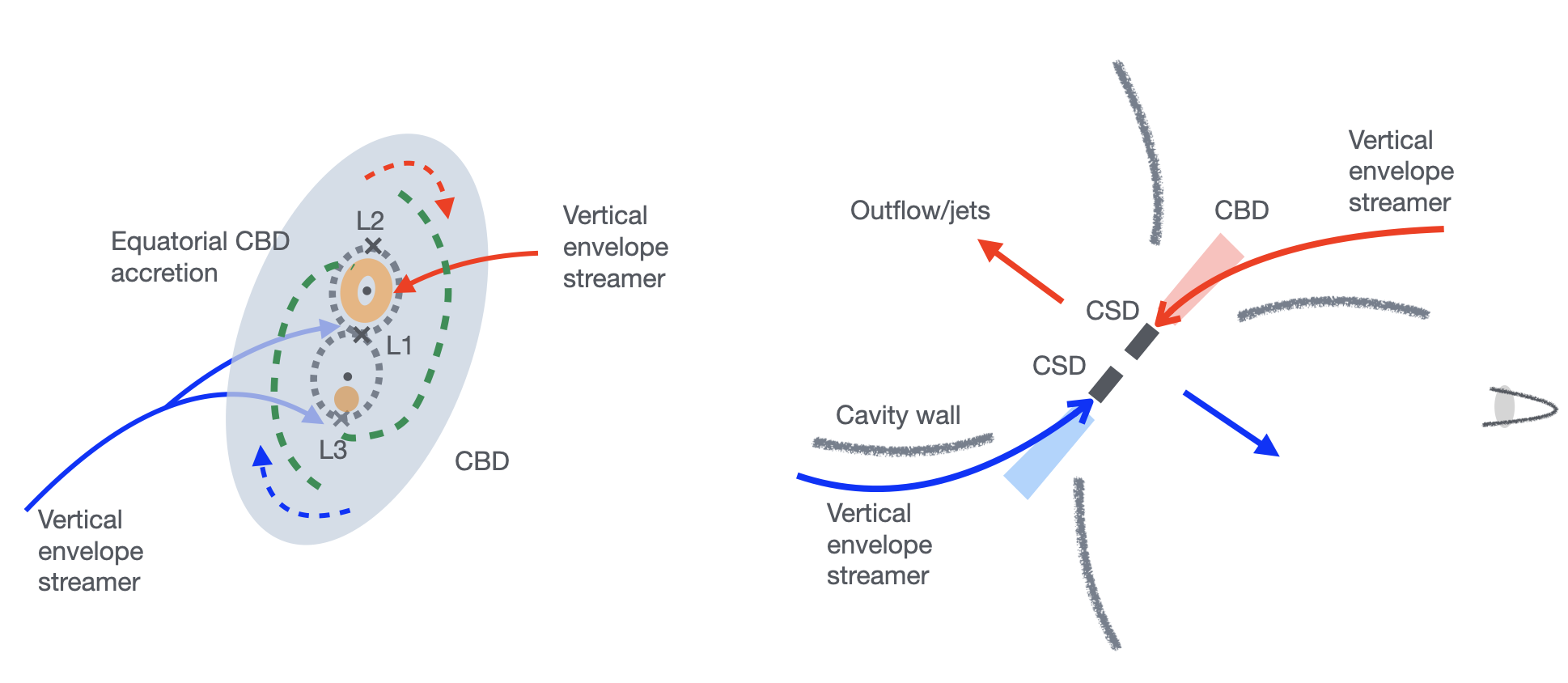}
    \caption{Schematic view of the L1551 IRS 5 system. 
    (Left) Observer's view. The gray ellipse and the blue and red dashed arrows indicate the dust continuum CBD and its rotation, respectively. Crosses mark the Lagrangian points ($L_{1}$, $L_{2}$, and $L_{3}$). Within the Roche lobes (gray dotted boundaries), orange regions denote COM emission around the northern and southern CSDs. Green dashed lines trace the continuum spiral arms; outside the Roche lobes these arms represent density enhancements in the CBD, while inside they transition into accretion flows feeding the CSDs.
    Solid blue and red arrows represent blue- and red-shifted SO streamers tracing vertical envelope-to-disk impacts onto the CSDs; where the blue streamer passes behind the CBD midplane, the arrows are shown with less opacity.
    (Right) LOS structure viewed from the right. The CSDs and rotating CBD are depicted as dark gray and red and blue rectangles, respectively. Straight blue and red arrows show the orientation of the bipolar outflow and jets bounded by cavity walls (gray curves). The vertical SO streamers are indicated by curved solid lines.
    \label{fig:schematic_L1551}}
\end{figure*}

For Source N, the observed spatial and kinematic distributions of molecules suggest a ``dual-mode accretion'' scenario.
In addition to the thermal heating from the central source due to the FUor event, two distinct feeding pathways coexist and induce different chemical responses in the surrounding environment, as illustrated in the schematic view in Figure~\ref{fig:schematic_L1551}.

Steady Equatorial CBD-to-CSD Accretion (North):
The northern enhancement of COMs spatially coincides with the spiral arm root predicted by CBD-to-CSD accretion simulations \citep{Takakuwa:2020tg}.
This region is characterized by high column densities, traced by \ce{CH3OH} and a wide variety of COMs.
The kinematic continuity from the CBD, manifesting as a positive velocity residual in the northeastern region, suggests a ``steady equatorial accretion'' mode in which material gradually accumulates along the disk plane.
Given the positional association of the COM emission with the continuum spiral arm and its velocity offset from the Keplerian rotation, the COMs in this region are possibly associated with the steady accretion traced by the arm. In this picture, the observed COMs likely originate from a combination of moderate accretion shocks and thermal processing, which is consistent with the relatively weak \ce{SO} at the arm root (Sect.~\ref{sec:shockintensity}).

Vertical Envelope-to-Disk Streamer Impact (Southeast and West):
We observed multiple large-scale filamentary SO structures around the binary, as discussed in Sect.~3.3. Because SO can trace either shocks along outflow cavity walls or infalling streamers feeding the system, the morphology of the observed filaments alone does not uniquely distinguish between these possibilities \citep[e.g.,][]{Pineda:2020aa, Valdivia-Mena:2022aa, Lee:2023vl, Liu:2025aa}. In the following, we therefore focus only on the two filaments highlighted in the schematic view, whose kinematics are more consistent with infall. As these filaments approach the CSDs, their line-of-sight velocities reach or closely approach the extreme rotational velocities of the disks, and their velocity gradients tend to be steeper toward the sources, as traced by the $0.43$~km~s$^{-1}$ channel maps (Figures~\ref{fig:chan_so_in} and \ref{fig:chan_so_out}) and the position--velocity diagrams, in which such gradients appear along the southeastern branch toward $L_3$ and the western extension (Figure~\ref{fig:pv_se}, first and third panels). Such behavior is more naturally interpreted as infall toward the disks than as outflow shocks, and is broadly consistent with the kinematic features of large-scale SO streamers delivering material to protostellar disks \citep[e.g.,][]{Lee:2023vl, Liu:2025aa}.

Among these two infall-like filaments, the blue-shifted southeastern streamer appears to be directly related to the localized COM activity in both Source N and Source S. In the channel maps at $V_{\rm LSR}\simeq+6.3$ to $+5.0$~km~s$^{-1}$, the southeastern structure connects toward the nearside of Source~N around the $L_1$ region (Figure~\ref{fig:chan_so_in}), and it coincides in position and velocity with the high-excitation COM emission on the nearside (Figures~\ref{fig:chan_ch3ocho} and \ref{fig:chan_c2h5cn}), which extends well beyond the compact blue-shifted excess in the residual map (Sect.~\ref{sec:spectra}). At $V_{\rm LSR}\lesssim+4.5$~km~s$^{-1}$, the southeastern structure connects to the southern part of the Source~S CSD near $L_3$, where it again coincides in position and velocity with the compact \ce{CH3OH} and DCN emission and a localized \ce{SO} enhancement (Figure~\ref{fig:pv_se}). Since the thermal heating from the central sources cannot explain these localized high-excitation components (Sect.~4.1.1), the streamer impact releasing grain-surface species, as seen in other protostellar systems \citep[e.g.,][]{Lee:2023vl, Liu:2025aa}, is the most plausible origin of these COMs.

The selection of the affected transitions is set by the excitation conditions of the shocked gas. The southeastern quadrant traces a hot, dense post-shock layer in which high-excitation transitions are populated, whereas the cooler thermally desorbed reservoir on the farside rim emits predominantly in low-excitation lines. This naturally explains why the high-$E_{\rm up}$ transitions of \ce{CH3OCHO} and \ce{C2H5CN} appear on the nearside while their low-$E_{\rm up}$ counterparts follow the farside morphology. Formamide (\ce{HCONH2}, $E_{\rm up}\simeq79$~K), although low in excitation, groups with the nearside high-excitation species, consistent with its frequent association with shocked or hot gas \citep{Mendoza:2014aa, Lopez-Sepulcre:2019vh}.

In contrast, the most red-shifted SO streamer connected to the western side of Source N is unlikely to be the direct origin of the COM emission observed there. The brightest \ce{SO} peak in the system is morphologically associated with the western extensions of the COMs and \ce{SO}, indicating a connection between them. However, the \ce{SO} peak is clearly offset from the disk COM emission (see Fig.~\ref{fig:mom0} and the top panel of Figure~\ref{fig:radial}), which is dominated by low-excitation transitions and is more naturally explained by thermal heating (Sect.~4.1.1). This offset suggests that the western \ce{SO} stream is probably not the main origin of the farside disk COMs.

\subsection{Diagnosing shock properties via chemical comparisons}
\subsubsection{Source S: Origin of the chemical dichotomy}

An intriguing feature in Source S is the morphological contrast between the localized emission of methanol (\ce{CH3OH}) and the ring-like structure of \ce{SO}. Methanol is obscured at the center by the high dust opacity, as in Source~N (Sect.~4.1.1), and is visible only in the localized region near the $L_3$ impact point, while \ce{SO} emission traces a complete loop along the outer edge of the disk. We propose that this dichotomy arises from the interplay between the binary-regulated disk structure and the distinct chemical properties of these species, specifically their sublimation conditions and post-shock regeneration processes.

First, we address the physical origin of the \ce{SO} ring. In isolated protostellar systems, rings of shock tracers such as \ce{SO} are often interpreted as centrifugal barriers, where the centrifugal force balances gravity, causing infalling gas to pile up and create a shocked region \citep[e.g.,][]{Sakai:2014aa, Liu:2025aa}. However, in the L1551 IRS 5 binary system, the disk size is primarily regulated by binary interactions rather than angular momentum conservation alone. The observed radius of the \ce{SO} ring ($\sim 20$~au) is in excellent agreement with the theoretical tidal truncation radius ($R_{\text{tid}} \approx 0.3a \text{--} 0.4a$) predicted for such systems \citep{Artymowicz:1994aa, Harris:2012aa}. This suggests that the \ce{SO} ring delineates the physical boundary of the disk determined by tidal truncation.

The mass accretion onto Source S is highly anisotropic, with fresh material being injected exclusively through the $L_3$ point. Consequently, the observed ring structure is not a result of uniform infall along the disk perimeter; rather, it originates from the rotational transport of shocked gas injected at the $L_3$ impact site. The differing spatial distributions of methanol and \ce{SO} can thus be explained by their survival timescales as the gas rotates away from this high-energy impact zone.

Methanol has a relatively high binding energy and requires dust temperatures of $T_{\rm dust}\gtrsim100$~K to desorb and remain in the gas phase \citep{CollingsEtAl2004}.
As the shocked gas rotates away from L3, it rapidly cools, and gas-phase CH$_3$OH is then efficiently removed by re-adsorption onto dust grains. Its depletion (freeze-out) timescale is estimated to be $t_{\rm dep}\sim0.5$--$50$~yr, evaluated from the standard gas--grain adsorption rate \citep{Lee:2004aa}, $t_{\rm dep} \simeq (6\times10^9 / n_{\rm H_2})\sqrt{m/T_k}$~yr,
where $m$ is the molecular weight in atomic mass units and $T_{k}$ the gas kinetic temperature in K. We adopt $T_{k}=50$~K \citep{Liu:2025aa, Sakai:2014aa} at the $\sim$20~au location of the localized methanol peak, which lies beyond the $\sim$16~au water snowline ($T_{\rm dust}\sim100$~K) of Sect.~4.1.1, and densities typical of disk midplanes and rims on tens-of-au scales, $n_{\rm H_{2}}\sim10^{8}$--$10^{10}$~cm$^{-3}$ \citep[e.g.,][]{ArturDeLaVillarmois:2022aa,DutreyEtAl2014PPVI}. This is orders of magnitude shorter than the local rotational timescale ($t_{\rm rot}\sim200$~yr at 20~au, assuming Keplerian rotation with $M_\ast=0.25\,M_\odot$), naturally confining CH$_{3}$OH emission to the immediate vicinity of the L3 shock site.

In contrast, sulfur-bearing species such as \ce{SO} (and its precursors such as \ce{H2S} or \ce{OCS}) have lower binding energies and can remain in the gas phase at significantly lower temperatures ($T_{\text{dust}} \sim 40\text{--}60$~K) \citep{Wakelam_2017}. This allows \ce{SO} to survive in the cooler post-shock regions propagating along the tidal truncation boundary. Furthermore, unlike methanol, \ce{SO} is efficiently reformed in warm post-shock gas via neutral-neutral reactions, such as \ce{ S + OH -> SO + H}.
This reaction is highly efficient in shocked regions where OH abundance is enhanced \citep{Pineau-des-Forets:1993aa, van-Gelder:2021wj}. This gas-phase regeneration extends the lifetime of \ce{SO}, allowing it to act as a tracer of ``slow shocks'' (velocity jumps of a few km~s$^{-1}$) along the disk rim. Consequently, the \ce{SO} molecules are rotationally smeared into a complete ring structure, whereas methanol remains localized, effectively tracing only the high-energy injection point.

\subsubsection{Source N: Shock intensity and accretion pathways}\label{sec:shockintensity}
In Source N, a notable feature is the spatial variation of \ce{SO} intensity. The \ce{SO} emission is bright in the streamer-related accretion shock near the $L_1$ point and the western extension, whereas the spiral foot associated with steady accretion exhibits significantly weaker \ce{SO} emission (see the moment~0 map and the channel maps around $+9.8$ to $+12$~km~s$^{-1}$ in Figure~\ref{fig:chan_so_in}). We interpret this contrast in the context of shock chemistry dependent on impact velocity.

According to \citet{van-Gelder:2021wj}, the abundance of gas-phase \ce{SO} is highly sensitive to the shock velocity ($v_s$). In low-velocity shocks ($v_s \sim 4$ km~s$^{-1}$), the dust temperature remains insufficient to trigger thermal sublimation of \ce{SO} from grain surfaces. In contrast, stronger shocks with $v_s > 4$ km~s$^{-1}$ are hot enough to thermally sublimate \ce{SO} ice. Crucially, in these energetic shock regions, \ce{SO} abundance is further enhanced by efficient gas-phase formation (e.g., \ce{S + OH -> SO + H}), leading to the bright emission observed.
The streamer impact is expected to drive shocks well above this threshold, capable of both liberating sulfur-bearing species from grains and triggering gas-phase reformation chemistry. On the other hand, the spiral arm structure, formed by steady equatorial accretion, shows markedly weaker \ce{SO} emission. Within the same shock-velocity picture, this weaker \ce{SO} points to milder shocks that are below the threshold for efficient gas-phase \ce{SO} enhancement. This is consistent with the expectation that in-plane accretion along the spiral arms is gentler than the vertical infall of the streamers, although we do not have a direct shock-velocity measurement.

This distinction has important implications for the origin of COMs detected in the spiral region. Unlike the streamer region where molecules are liberated and processed by strong shocks, the COMs in the spiral foot are unlikely to be of shock origin given the weak \ce{SO} emission. Instead, we propose that these molecules trace the material piled up by steady accretion, which has been thermally desorbed by the elevated luminosity from the recent FU Orionis-type outburst of the central source. In this radiatively heated region, gas-phase \ce{SO} is limited to the sublimated ice content, lacking the significant abundance boost provided by high-temperature shock reformation seen in the streamer.

\section{Conclusions} \label{sec:conclusion}

In this study, we have presented high-resolution ALMA observations of COMs and SO toward the Class I binary protostellar system L1551 IRS 5. By analyzing the spatial and kinematic distributions of these species, we have disentangled the thermal and dynamic history of the system. Our main observational findings and their physical implications are summarized below.

\begin{enumerate}
    \item Radiative Heating from the Central Outburst: We observed a widespread distribution of low-excitation COMs filling the Roche lobe of Source N, whereas COM emission in Source S is highly compact and offset from the center. This indicates that the recent FUor-type outburst in Source N globally sublimated ice mantles within its own vicinity, while Source S remains largely shielded or unaffected by this global heating event.

    \item Dual-Mode Accretion and Streamer-Foot Shocks: We identified kinematically and morphologically distinct accretion flows feeding the protostellar disks. 
    In Source N, COMs along the northeastern spiral arm are consistent with a steady, equatorial CBD-to-CSD accretion flow.
    In contrast, large-scale ($\sim$600~au) SO streamers are connected to the localized high-excitation COM emission along the eastern (nearside) disk surface of Source~N (near $L_1$) and to the compact COM emission in Source~S (near $L_3$).
    These spatial and kinematic connections suggest that infalling streamers deliver material directly from the envelope to the CSDs and induce localized streamer-foot shocks at their landing sites in both sources.

    \item Chemical Diagnostics of Shock Properties:
    The spatial contrast between SO and methanol distributions serves as a powerful diagnostic tool for shock properties and post-shock evolution. In Source N, strong SO emission at the streamer impact site indicates high-velocity shocks capable of triggering gas-phase SO reformation, distinguishing it from the weak SO emission in the steady spiral inflow. In Source S, a localized methanol spot coexists with a ring-like SO structure. This dichotomy reflects differences in post-shock survival timescales; methanol rapidly freezes out as the gas rotates and cools, whereas SO is efficiently regenerated in the gas phase and rotationally smeared along the tidal truncation boundary.
\end{enumerate}

This study highlights the utility of combining COMs and simple shock tracers such as SO to spatially separate material accumulated via steady spiral inflows from that injected via high-velocity streamer impacts. L1551 IRS 5 stands out as an ideal laboratory for studying these shock-induced chemical processes during the active phase of binary star formation.

\begin{acknowledgements}
This paper makes use of the following ALMA data: ADS/JAO.ALMA\#2016.1.00209.S and ADS/JAO.ALMA\#2016.1.00138.S.
ALMA is a partnership of ESO (representing its member states), NSF (USA) and NINS (Japan), together with NRC (Canada), MOST and ASIAA (Taiwan), and KASI (Republic of Korea), in cooperation with the Republic of Chile. The Joint ALMA Observatory is operated by ESO, AUI/NRAO and NAOJ.
S.T. is supported by JSPS KAKENHI grant Nos. JP21H00048, JP21H04495, and JP24K00674, and by NAOJ ALMA Scientific Research grant No. 2022-20A.
J.-E. Lee was supported by the National Research Foundation of Korea (NRF) grant funded by the Korea government (MSIT) (grant numbers RS-2024-00416859 and RS-2026-25490557).
This research was supported by the Korea Astronomy and Space Science Institute under the R\&D program (Project No. 2026-1-844-00) supervised by the Korea AeroSpace Administration.
\end{acknowledgements}

\bibliographystyle{aa}
\bibliography{ref}

\clearpage
\onecolumn
\nolinenumbers
\begin{appendix}
\section{Moment~0 maps of individual molecular transitions}\label{app:commom0}
\begin{figure}[h!]
\centering
\includegraphics[angle=0,width=180mm]{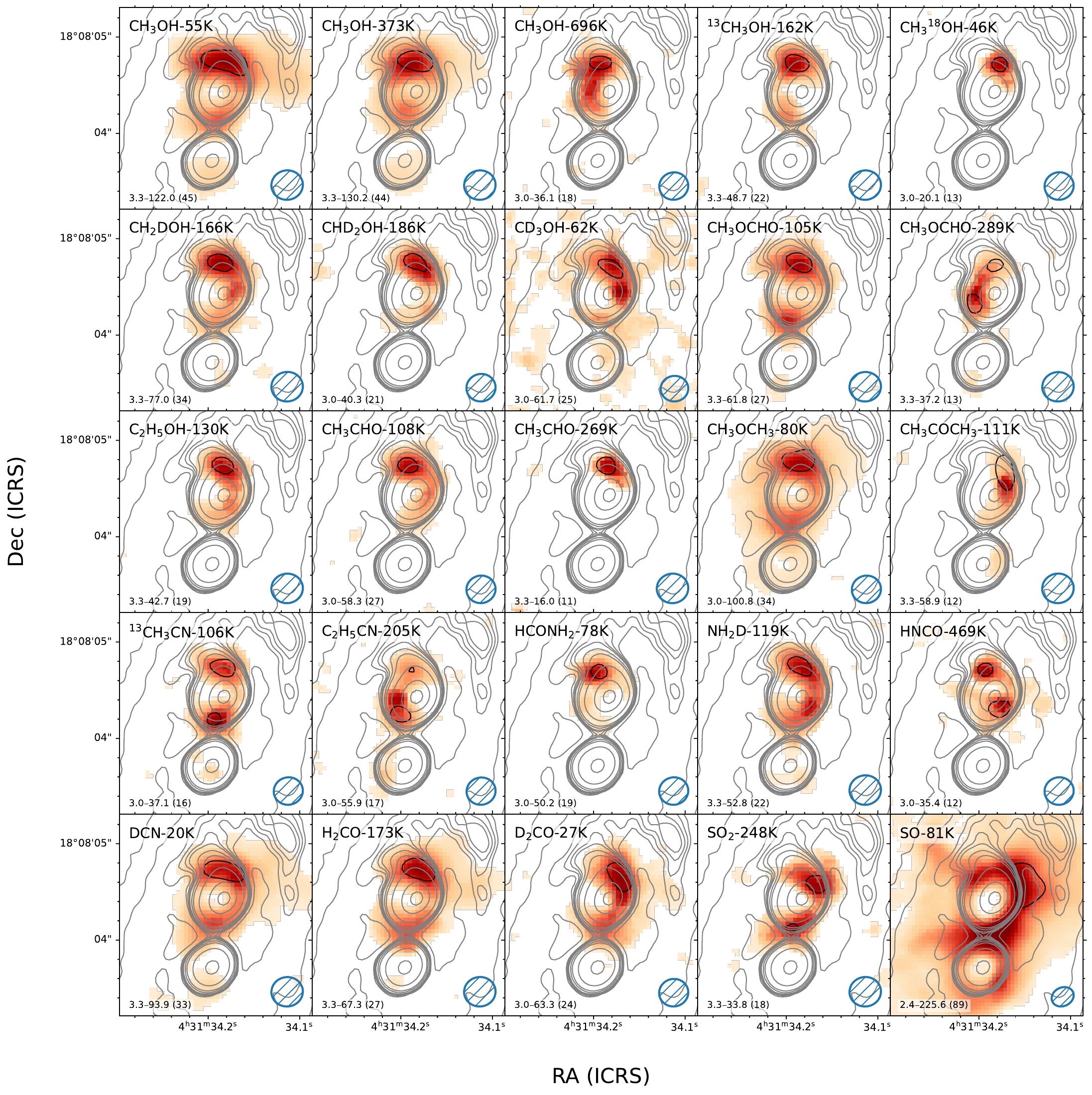}
\caption{Integrated intensity (moment 0) maps of molecular lines toward Source~N,
overlaid with dust continuum contours.
The adopted velocity range is $2 < v_{\rm LSR} < 11$~km~s$^{-1}$,
except for NH$_2$D, which uses $3.6 < v_{\rm LSR} < 11$~km~s$^{-1}$
to avoid contamination from a neighboring line.
The molecule name and upper-level energy are indicated at the top left of each panel.
The gray contours are the same as in Fig.~\ref{fig:anticorr}.
The black contours denote 80\% of the peak intensity (moment~8).
The minimum and maximum values of each moment~0 map, followed by the peak intensity of the moment~8 map in parentheses, are shown at the bottom left of each panel in units of Jy~beam$^{-1}$~km~s$^{-1}$ and mJy~beam$^{-1}$, respectively.
\label{fig:mom0}}
\end{figure}

Figure~\ref{fig:mom0} presents the integrated intensity (moment~0) maps of the individual molecular transitions identified in Figure~\ref{fig:spec0}. The \ce{SO} map is an exception. It is not drawn from the spectral windows used for the COM identification but from the separate \ce{SO} observations with higher velocity resolution described in Sect.~\ref{sec:data}.

\clearpage
\section{Supplementary \ce{SO} channel maps}\label{app:chan}
\begin{figure}[h!]
\centering
\includegraphics[width=180mm]{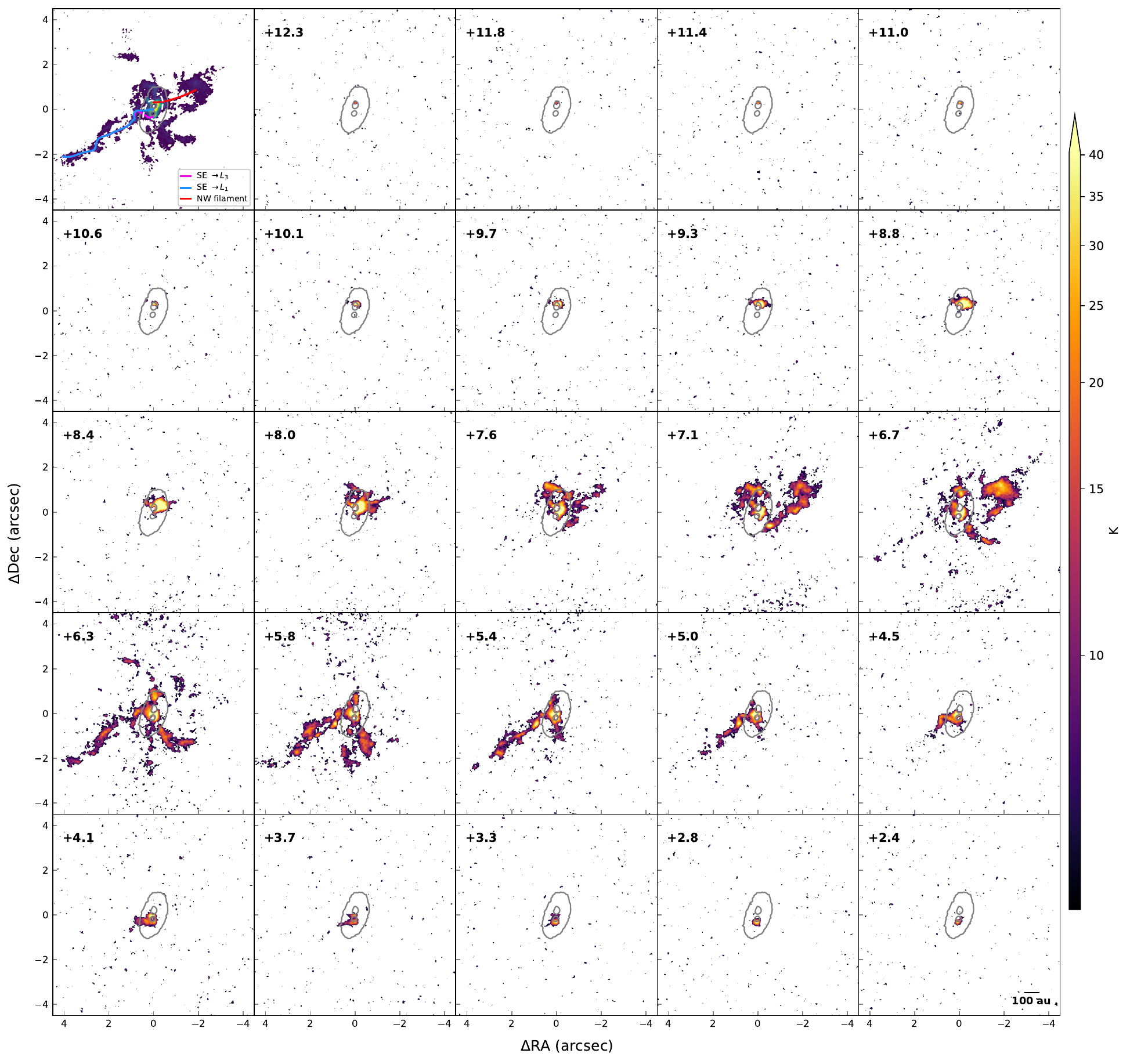}
\caption{Large-scale \ce{SO} channel maps at the same velocity sampling as Figure~\ref{fig:chan_so_in}, with the same three streamer paths overlaid in the first panel. The color scale shows the brightness temperature (K), and the LSR velocity (km~s$^{-1}$) is given in the upper left of each panel. Gray contours show the 0.9~mm continuum at 6 and $50\sigma_{\rm cont}$, whose inner contours mark the two protostars; a 100~au scale bar is shown in the last panel.\label{fig:chan_so_out}}
\end{figure}

Figure~\ref{fig:chan_so_out} shows the \ce{SO} channel maps over an extended field of view, with the same velocity sampling ($0.43$~km~s$^{-1}$) as the zoomed-in maps in Figure~\ref{fig:chan_so_in}. The full extents of the southeastern strand and of the northwestern filament, and their connections to the structures seen in the zoomed-in maps, are visible.

\begin{figure}[h!]
\centering
\includegraphics[width=180mm]{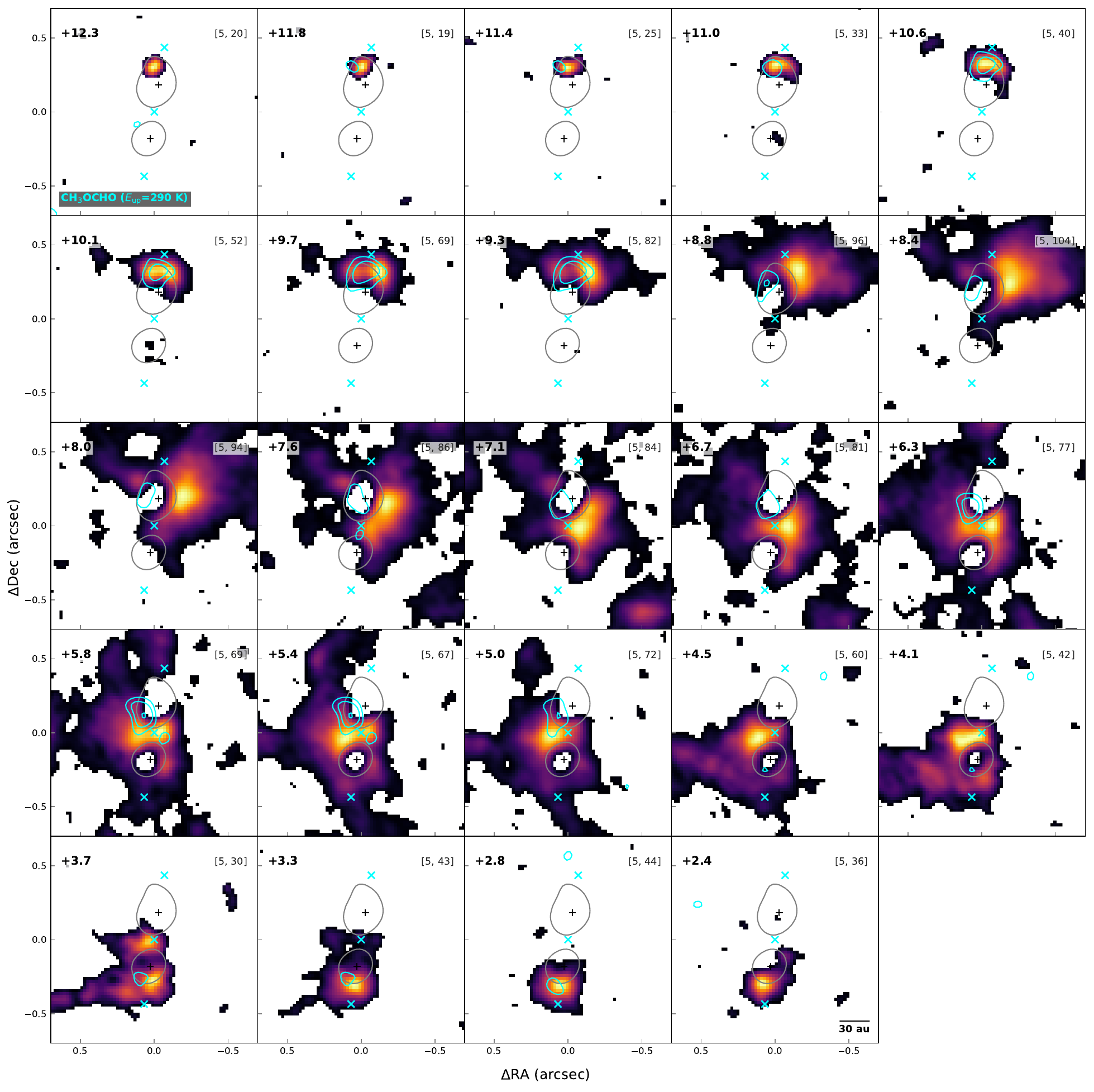}
\caption{Zoomed-in \ce{SO} channel maps overlaid with cyan contours of \ce{CH3OCHO} ($E_{\rm up}=290$~K) at 3, 5, 8, and 12 times the rms noise, where the rms of 1.6~mJy~beam$^{-1}$ is measured in emission-free channels. Markers and continuum contours are the same as in Figure~\ref{fig:chan_so_in}.\label{fig:chan_ch3ocho}}
\end{figure}

\begin{figure}[h!]
\centering
\includegraphics[width=180mm]{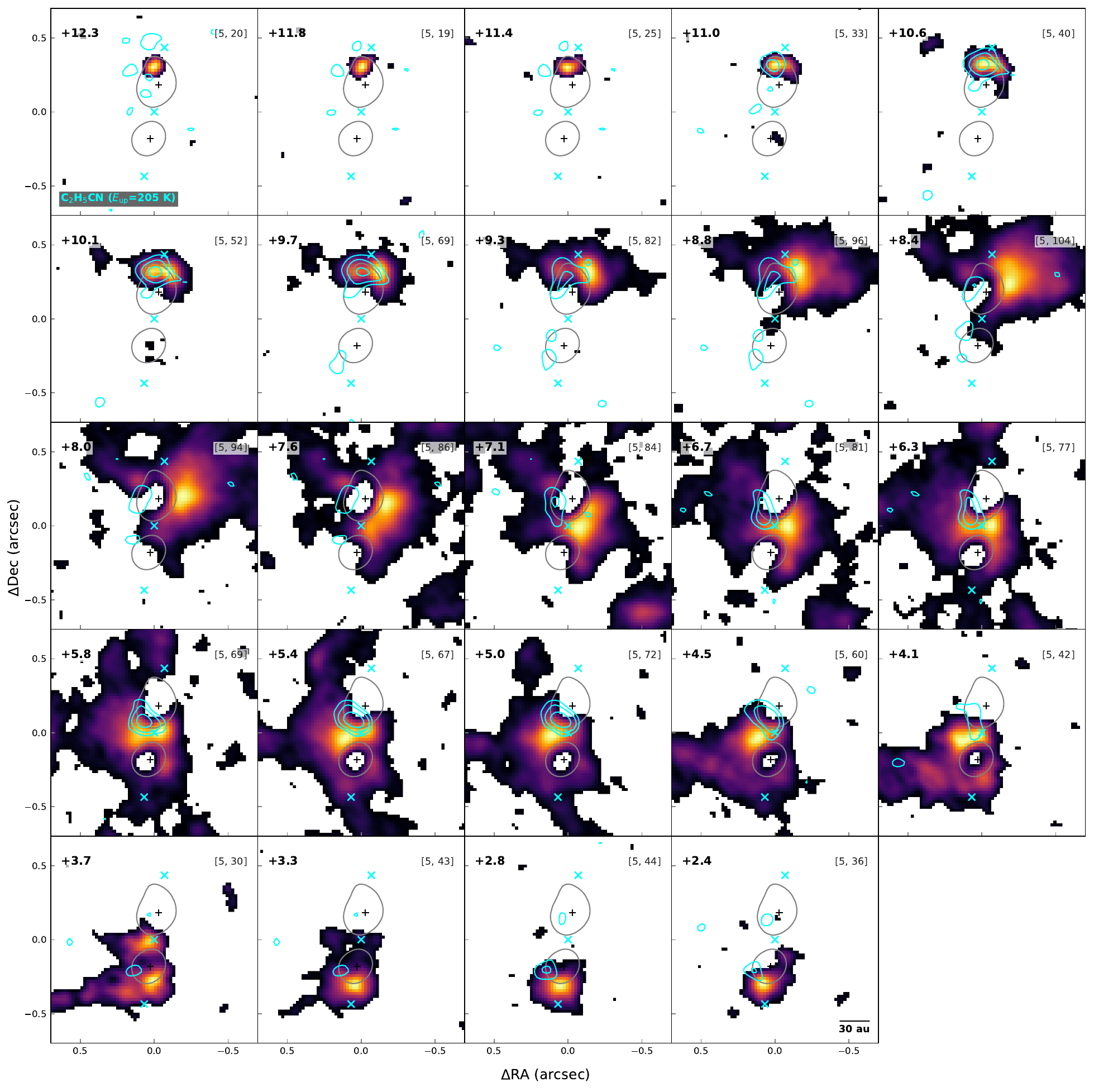}
\caption{Same as Figure~\ref{fig:chan_ch3ocho}, but for \ce{C2H5CN} ($E_{\rm up}=205$~K) with an rms of 1.6~mJy~beam$^{-1}$.\label{fig:chan_c2h5cn}}
\end{figure}

\clearpage
\nolinenumbers
Figures~\ref{fig:chan_ch3ocho} and \ref{fig:chan_c2h5cn} compare the zoomed-in \ce{SO} channel maps with the channel-map contours of two representative high-excitation COM transitions, \ce{CH3OCHO} ($E_{\rm up}=290$~K) and \ce{C2H5CN} ($E_{\rm up}=205$~K). Both molecules show the blue-shifted component at $V_{\rm LSR}\simeq+6$~km~s$^{-1}$ on the eastern disk surface of Source~N, near the region where the southeastern \ce{SO} branch approaches $L_1$, and the red-shifted component at $V_{\rm LSR}\simeq+8$ to $+10$~km~s$^{-1}$ at the spiral-arm root north of Source~N. \ce{CH3OH} ($E_{\rm up}=696$~K) shows a consistent pattern.

\clearpage
\end{appendix}

\end{document}